\documentclass[useAMS]{mn2e}
\usepackage{mnras_cite}
\usepackage{epsfig}
\usepackage{deluxetable}
\usepackage{subfigure}

\begin{document}
\title[A mid-IR study of Lyman Break Galaxies]{On the Stellar Masses of IRAC detected Lyman Break Galaxies at $z$ $\sim$ 3}\author[Magdis et. al]{G.E. Magdis$^{1,2}$, D. Rigopoulou$^{1,3}$, J.-S. Huang$^{4}$, G.G. Fazio$^{4}$\\
\vspace{0.5cm}\\
$^{1}$Department of Physics, University of Oxford, Keble Road, Oxford OX1 3RH, United Kingdom\\
$^{2}$CEA Saclay/Service d'Astrophysique, CNRS F-91191 Gif-sur-Yvette Cedex, France \\
$^{3}$Rutherford Appleton Laboratory, Chilton, Didcot, OX11 0QX, United Kingdom\\
$^{4}$Harvard-Smithsonian Center for Astrophysics, 60 Garden Street, Cambridge, MA 02138\\ }

\maketitle

\begin{abstract}

We present results of a large survey of the mid--IR properties of
248 Lyman Break Galaxies with confirmed spectroscopic redshift using
deep Spitzer/IRAC observations in six cosmological fields. By
combining the new mid--IR photometry with optical and near-infrared
observations we model the Spectral Energy Distributions (SEDs)
employing a revised version of the Bruzual and Charlot synthesis
population code that incorporates a new treatment of the TP--AGB phase
(CB07). Our primary aim is to investigate the impact of
the AGB phase in the stellar masses of the LBGs, and compare our new results
with previous stellar mass estimates. We investigate the stellar mass of
the LBG population as a whole and assess the benefits of adding longer 
wavelengths to estimates of stellar masses for high redshift galaxies. 
Based on the new CB07 code we find
that the stellar masses of LBGs are smaller on
average by a factor of $\sim$ 1.4 compared to previous estimates.
LBGs with 8$\mu$m and$/$or 24$\mu$m detections show higher masses
(M$_\ast$ $\sim$ 10$^{11}$ M$_\odot$)  than LBGs faint in the IRAC bands
(M$_\ast$ $\sim$ 10$^{9}$ M$_\odot$).
The ages of these massive LBGs are considerably higher than the rest of
the population, indicating that they have been star-forming for at least 
$\sim$ 1 Gyr. We also show how the addition of the IRAC bands, improves the
accuracy of the estimated stellar masses and reduced the scatter on
the derived M/L ratios. In particular, we present a tight
correlation between the 8$\mu$m IRAC band (rest--frame K for galaxies
at z$\sim$3) and the stellar mass.  We calculate the number density
of massive (M$_\ast$ $>$ 10$^{11}$ M$_\odot$) LBGs and find it to be
$\Phi$= (1.12 $\pm$ 0.4) $\times$ 10$^{-5}$ Mpc$^{-3}$, $\sim$1.5 times lower than that found by previous studies.
Finally, based on UV-corrected SFRs we
investigate the SFR-stellar mass correlation at $z$ $\sim$ 3, find it
similar to the one observed at other redshifts and show that our data
place the peak of the evolution of the specific star formation rate at
$z$ $\sim$ 3.

\end{abstract}

\section{Introduction}

Although there has been considerable progress in understanding galaxy
formation and evolution, current theoretical models still suffer from
several uncertainties. These uncertainties are mainly introduced by
parameters that are very difficult to constraint, such as the initial
mass function (IMF), the action of feedback by supernovae and stellar
winds and the chemical evolution. Given that theoretical guidance is
so uncertain, direct empirical information is essential in driving the
investigation. The detection and study of high--$z$ galaxies can put
constraints on the physical parameters, put the existing models to the
test and guide us to a better and more accurate description of the
early galaxies.  To this end, recent surveys have come to play a
central role in modern cosmology, revealing a wealth of $z$ $\sim$
2--3 star--forming galaxies.

Among others, there are two efficient methods of detecting high--$z$
galaxies. The first relies on sub-millimetre blank field observations
using the Sub-millimetre Common-User Bolometer Array (SCUBA) on the
James Clerk Maxwell Telescope (e.g. Hughes et al. 1998) or the Max
Plank Millimetre Bolometer array (MAMBO, e.g. Bertoldi et al. 2000)
revealing the population of the sub-millimetre galaxies (SMGs) at $z$
$>$ 2 (e.g. Chapman et al. 2000; Ivison et al. 2002; Smail et
al. 2002). The second relies on colour selection criteria, selecting
high--$z$ galaxies with different characteristics. For instance, the
BzK colour criterion introduced by Daddi et al. (2003) selects
moderately aged and moderately obscured star forming galaxies at $z$
$\sim$2 while the J$_{s}$ $-$ K$_{s}$ $>$ 2.3 criterion (Franx et
al. 2003; van Dokkum et al. 2003) detects both strongly obscured by
dust high--$z$ star--forming galaxies as well as massive/evolved
systems. One of the most common methods in the last decade though, now
comprising an impressive catalogue of thousands of galaxies at $z$
$\sim$ 3, has been the selection based on photometric redshift gained
from observations of the ``Lyman Break'' located at 912{\AA} in the
spectrum of a star--forming galaxy.  The Lyman Break Galaxy (LBG)
selection method was pioneered by Steidel et al. (1996, 1999, 2000,
2003), but has also been used by many other groups (e.g. Madau et
al. 1996; Pettini et al. 2001; Bunker et al. 2004; Stanway et
al. 2004; Ouchi et al. 2004a,b).

Since their detection, LBGs have caught the attention of the
scientific community, and a considerable amount of effort has been
concentrated on investigating and understanding their nature. To this
direction, several strategies have been employed, ranging from
multi--wavelength photometric observations (from X--rays (Nandra et
al. 2002; Laird et al. 2006), to sub-mm (Chapman et al. 2000; Ivison
et al. 2005)), to optical and near--infrared spectroscopy (e.g. Erb et
al. 2003; Pettini et al. 2001; Steidel et al. 1996a; Shapley et
al. 2003).

One way to derive the properties of the LBGs is by fitting the
observed SED with model SEDs, generated by stellar synthesis
population codes. This technique was first applied by Sawicki \& Yee
(1998) to a sample of 17 LBGs from the HDF-N (Williams et
al. 1996). Later it was further employed by Papovich et al. 2001 and
Shapley et al. 2001 with NIR (rest--frame optical) photometry.  Both
groups derive similar stellar masses $\sim$10$^{10}$M$_\odot$ and
agree that the stellar mass is generally well constrained by the
fitting procedure in contrast to star formation rate and stellar age
which suffer from large uncertainties. Any study based only on
rest--frame UV/optical data though, is far from complete, as the UV
light of these galaxies samples only the short--lived massive stars of
the forming populations. For a more comprehensive study, rest--frame
near--IR observations of galaxies are essential since they trace the
bulk of the stellar emission. The advent of Spitzer Space Telescope,
and its unprecedented sensitivity opened a window to the IR part of
the spectrum of high galaxies and enabled the study of their infrared
properties.

Up to date, there have been few mid--IR studies of LBGs. Using Spitzer
observations, Barmby et al. (2004) investigated the properties of
several hundred LBGs in Q1700, while Reddy et al. (2006), based on a
sample of UV selected galaxies in HDFN, found that they span 2 orders
of magnitude in age and stellar mass and 4 orders of magnitude in dust
obscuration.  More recently, Shim et al. (2007) presented a study
based on a sample of LBGs from a sub--region of the First Look
Survey. Although their sample lacks spectroscopic redshift they find
that more than $\sim$70\% of the IRAC detected LBGs have estimated
stellar masses $>$10$^{11}$M$_\odot$. A detailed mid--IR study of LBGs
has been presented by Rigopoulou et al. (2006) (R06 hereafter). Based
on a sample of 175 LBGs with confirmed spectroscopic redshift in the
Extended Groth Strip they suggested that LBGs with bright IRAC colours
are more massive (M$_\ast$$\sim$10$^{11}$M$_\odot$), older (t$_{sf}$
$\sim$ 1Gyr) and suffer more extinction when compared to the rest of
the sample. Magdis et al. (2008), based on a a sample of 751 LBGs,
presented the mid--IR colours of LBGs and suggested that they are a
rather inhomogeneous population ranging from those that have bright
IRAC colours with SEDs that rise steeply towards longer wavelengths
and R $-$ [3.6] $>$ 1.5 (``red'' LBGs) to those that are faint in the
IRAC bands with R $-$ [3.6] $<$ 1.5 (``blue'' LBGs). Finally, Huang et
al. (2005),
 reported the detection of several LBGs in the Extended Groth Strip
 (EGS) at MIPS 24$\mu$m, revealing a subset of LBGs, the Infrared
 Luminous Lyman Break Galaxies (ILLBGs).

Despite those efforts, the mid--IR properties of the LBGs are not
clear yet. Partially, this is due that fact that the derived
properties of the population depend critically on the adopted stellar
synthesis population model. New developments of the stellar synthesis
population codes, incorporating more accurate prescriptions of the
stellar evolution phases and more particularly of the
Thermal-Pulsating Asymptotic Giant Branch phase (TP-AGB), allow us to
derive more realistic and robust estimates of the properties of the
population.

In this paper, we present mid--IR photometry for a sample of
spectroscopically confirmed $z$ $\sim$ 3 LBGs detected as part of the
IRAC Guaranteed Time Observations (GTO) program on 6 cosmological
fields. Benefited from ground based optical and IRAC observations, we
are in the unique position to constrain the properties of the
population. Although the stellar masses of the LBGs have been
studied in the past, here we make use of a revised version of the
Bruzual and Charlot stellar synthesis population code (CB07, private
communication) that incorporates a new treatment of the TP--AGB phase
and extend our study to explore the SFR-stellar mass relation at
z$=$3.  Our aims are to investigate the impact of the new AGB--phase
recipe on the stellar masses of the population, derive the range of
the stellar masses of the LBGs and assess the benefits of adding
longer wavelengths. In Section 2 we present a brief account of the
observations and data reduction while in Section 3 we describe the
model parameters and the SED fitting of the observed SEDs, used to
derive the properties of the population. Section 4 is dedicated to
derived stellar masses. We quantify the impact of the addition of the
AGB phase in the model SEDs, discuss the range of stellar masses of
our sample and present the M/L as a function of wavelength. In Section
5 we focus on the number density of massive LBGs and present the
stellar mass density of our sample. Finally, in Section 6 we
investigate the SFR-stellar mass correlation for star-forming galaxies
at $z$=3 while in Section 7 we summarise our results.

\section{The Spitzer LBG sample}

The data for this study have been obtained with the Infrared Array
Camera (IRAC) (Fazio et al. 2004) on board the Spitzer Space
Telescope. The majority of our data are part of the IRAC Guaranteed
Time Observation program (GTO, PI G. Fazio) and include the fields:
Q1422+2309 (Q1422), DSF2237a,b (DSF), Q2233+1341 (Q2233), SSA22a,b
(SSA22) and B20902+34 (B0902) while data for the HDFN come from the
Great Observatories Origin Deep Survey program (GOODS, PI
M. Dickinson). The data analysis and the mid--IR identification of the
LBGs in these fields have been presented in detail by Magdis et
al. (2008). Based on a catalogue of 1261 LBGs by Steidel et al. (2003)
in these fields, we constructed a mid--IR sample of 751 LBGs, that
were observed in at least one IRAC band. The sample consists of three
categories of objects: those that have confirmed spectroscopic
redshift (through follow up ground-based optical/NIR spectroscopy,
Steidel et al. 2003) and are identified as galaxies at $z$ $\sim$ 3
(LBGs-z) or classified as active galactic nuclei (AGN)/QSO and those
that do not have spectroscopic redshift. In total, 321 LBGs-$z$, 12
AGN/QSO and 435 LBGs without spectroscopic redshift are covered. In
Table 1 we list the ground based and IRAC photometry of LBGs-$z$ from
the current sample, that are observed in all IRAC bands.

In the current study, we focus on a sub-sample of 186 LBGs that 1)are
observed in all IRAC bands, 2) are detected in at least one IRAC band,
3) have confirmed spectroscopic redshift and 4) are classified as
galaxies, lacking a signature of strong AGN activity in their rest--UV
spectrum (Shapley et al. 2003). In our sample we add 10 LBGs that
satisfy our criteria from the field Q1700.

Out of these, 71 LBGs are detected at 8$\mu$m, consisting the 8$\mu$m
sample. As discussed by Magdis et al. (2008), whether a LBG is
detected at 8$\mu$m relies critically both on the depth of the
observation and on how bright the LBG is in shorter wavelengths
(i.e. 3.6- and 4.5-$\mu$m). Since in HDFN, the deepest field of their
study, all LBGs with [3.6]$_{AB}$ $<$ 23 were detected at 8$\mu$m,
they extend the original 8$\mu$m sample, by including the LBGs of
shallower fields that were not detected at 8$\mu$m but with
[3.6]$_{AB}$ $<$ 23.  We will refer to this as the ``extended 8$\mu$m
sample'' and it consists of 105 LBGs. We also match our sample to the
MIPS 24$\mu$m catalogue of detected LBGs in HDFN published by Reddy et
al. (2006) (3$\sigma$,f$_{24}$ = 8$\mu$Jy). Among the 8$\mu$m detected
LBGs in HDFN, 18 LBGs are detected at 24$\mu$m, while one of our
galaxies (HDFN-M18), has f$_{24}$ = 88.2$\mu$Jy and according to Huang
et al. (2005) criterion is classified as a ILLBG (f$_{24}$ $ > $
60$\mu$Jy).

Although the current sample consists of LBGs lacking strong emission
lines in their optical spectrum, one cannot rule out the presence of
an obscured AGN. This issue was discussed by Magdis et al. (2008),
where they showed that the mid--IR colours of the LBGs are consistent
with that of star-forming galaxies and that AGNs at $z$ $\sim$ 3 tend
to have brighter 8$\mu$m fluxes and exhibit redder [4.5]-[8.0]
colours. Another way to explore this possibility would be to search
for object whose mid-IR SED is well fitted with a power law
($f$$_{\nu}$ $\propto$ $\nu$$^{\alpha}$ with $\alpha$ $\leq$ -0.5,
Donley et al. 2007, Alonso-Herrero et al. 2006), indicative of the
presence of hot dust heated by an AGN. However, the power law galaxy
(PLG) selection using IRAC photometric points, is not applicable at
the redshift range of our sample.  At $z$ $\sim$ 3 the IRAC bands
sample the blue part of the 1.6$\mu$m bump, dominated by light from
the stellar component of a galaxy. Hence, even if an AGN exists, one
cannot trace its signature at the wavelength range of 0.9-2$\mu$m
rest--frame. This has also been pointed out by Donley et al. (2008),
where they showed that at $z$ $>$2.5 the star-forming galaxy templates
have IRAC SEDs that meet the typical PLG criteria.  We note however,
that according to Table 5 of Magdis et al. (2008), the fraction of AGN
among the 8$\mu$m detected LBGs is $\sim$ 11$\%$, compared to $\sim$
3.5$\%$ among the whole sample of R $<$ 25.5 LBGs.  Finally, the mean
redshift of our sample is 2.92 (the distribution has $\sigma$ $=$
0.12).

\section{Derivation of Physical Properties}

We derive stellar masses for the LBG sample by fitting the latest CB07
stellar population synthesis models to the observed SEDs. The most
important update of CB07 when compared to its predecessor (BC03) is
the use of variable molecular opacities instead of the scaled--solar
tables. Also, the calibration is not only based on the reproduction of
the CSLFs in both Magellanic Clouds, but also on the data for C-- and
M--star counts in Magellanic Cloud clusters, providing the right
luminosities of the TP--AGB phase and the right contribution of
TP--AGB stars to the integrated light (Marigo \& Girardi 2007). In
this section, we first describe the CB07 models and the fitting
process that we use to infer the stellar masses and then comment on
the uncertainties introduced by this type of analysis.

\subsection{SED fitting.}

We use the CB07 code to generate model SEDs in order to fit the
observed SEDs of 196 LBGs with confirmed spectroscopic redshift and at
least one IRAC detection. Our aim is to derive the properties of the
population and more particularly the stellar masses. We adopted the
Padova 1994 stellar evolution tracks and constructed models with solar
metallicity (see discussion in Shapley et al. 2004) and a Chabrier
Initial Mass Function.  We use the Calzetti et al.  (2000) starburst
attenuation law to simulate the extinction by dust. We have considered
two simple single--component models: exponentially declining models of
the form SFR(t)$\propto$ exp(-t/$\tau$) with e-folding times of $\tau$
= 0.05, 0.1, 0.5, 1.0, 1.5, 2.0, and 5.0 Gyr and continuous star
formation (CSF) models ($\tau$ = $\infty$).

Our model fitting followed standard procedures applied in similar
studies of high-redshift galaxies (e.g. Papovich et al. 2001; Shapley
et al. 2001; Forster Schreiber et al. 2004; Shapley et al. 2005;
Papovich et al. 2006; Yan et al. 2006a). The free parameters of our
models are: dust extinction (E(B$-$V)), age (t$_{sf}$ defining the
onset of star formation), stellar mass ($M_{\ast}$ ), star formation (SFR) and star
formation history ($\tau$ ). For each of the star formation histories,
we generated models with ages ranging from 1 Myr to the age of the
Universe at the redshift of the galaxy being modeled, while we allowed
the dust extinction to vary between E(B $-$ V) = 0 and E(B $-$ V) =
1.0.  Furthermore, we computed the absorption by the intergalactic
medium of neutral hydrogen (Madau 1995) at the redshift of each
galaxy, and attenuated appropriately the SED of the generated
models. The model SEDs were then placed at the redshift of each galaxy
and were compared to the observed SEDs by computing a reduced
$\chi$$^{2}$. The CB07 spectra are normalized to an SFR of 1
M$_{\odot}$ yr$^{-1}$ for the continuous star formation model while
for the exponentially decaying models, the galaxy mass is normalized
to 1 M$_{\odot}$ as t=$\infty$. For each individual galaxy,
best-fitting parameters (age, E(B-V), and normalizations) were derived
from minimization of the reduced $\chi$$^2$. This normalizations were
then converted to best-fit stellar mass and along with the best fit
age, E(B-V) and $\tau$ was considered the overall ``best-fit''.
 
To quantify the error in the derived stellar masses, we compute the
range of normalizations that result in an SED with reduced
$\chi$$^{2}$ values within $\Delta$$\chi$$^{2}$$_{reduced}$ = 1 of the
minimum value. We adopt these as the 1$\sigma$ uncertainties
associated with the stellar masses. Finally, to facilitate comparisons
with previous studies in the fields as well as to quantify the impact
of the AGB phase on the derived stellar masses we choose to repeat our
analysis using this time models generated with BC03.

\subsection{Model Parameters and systematic uncertainties}

Of course, the derived best--fit parameters are subject to the adopted
parameters of the model fits. In what follows we discuss how varying
those properties affects the best--fit SED parameters and
uncertainties involved in this kind of analysis.

\subsubsection{Extinction}

The impact of different extinction laws has already been investigated
by e.g., Papovich et al. (2001), Dickinson et al. (2003), who found
the effect to be overall small. For the present work we have adopted
the Calzetti (2000) law. Such a law reproduces the total SFR from the
observed UV for the vast majority of LBGs and accurately predicts the
average X--ray and radio continuum fluxes of $z$ $\sim$ 2
star--forming galaxies.  (e.g. Reddy \& Steidel 2004, Reddy et
al. 2005, Nandra et al. 2003, Daddi et al. 2007).  The choice of the
Calzetti law was also dictated by the desire to facilitate comparison
with previous works in the field.

\subsubsection{Metallicity and Initial Mass Function}

So far, information on element abundances in LBGs is rather
limited. Pettini et al. (2002) determined element abundances in cB58,
a typical $L_{\ast}$ galaxy which benefits from a factor of 30
magnification, and found it to be $\sim$ 0.25 Z$_{\odot}$. Nagamine et
al. (2001) suggested that near-solar metallicities are in fact common
in $z$ $\sim$ 3 galaxies with masses greater than $10^{10}$
M$_{\odot}$, which is broadly consistent with the results for
cB58. Erb et al. (2006) and Shapley et al. (2004) also argued for
solar metallicities for $z$ $\sim$ 3 LBGs. Based on these results we
used solar metallicity in the models.  Reducing metallicity to half
solar would decrease the derived masses by 10--20\% (Papovich et
al. 2001).

Finally, although the issue of the IMF that best describes the stellar
population of high--$z$ galaxies still remains open (e.g. Renzini
2005; van Dokkum 2007; Dave 2008), recent results favor the scenario
of a top--heavy IMF (e.g. Baugh et al. (2005), Nagashima et
al. (2005), Lacey et al. (2008)).  Driven from these results we chose
to use a Chabrier IMF to generate the CB07 models.

\subsubsection{Systematic Uncertainties}

The systematic uncertainties, inherent in population synthesis
modeling, have been studied and extensively described in the
literature (e.g.. Papovich et al. 2001, Shapley et al. 2001, Shapley
et al 2005).  The limitations of this technique mainly originate from
the fact that the models cannot fully constrain the star formation
history of high--$z$ galaxies. In brief, the strong dependancy of the
inferred extinction and age to the adopted star formation history
introduces large degeneracies and makes the determination of these
parameters highly uncertain.

These uncertainties are of course likely to affect the inferred
stellar masses but in a less dramatic way. To test the impact of the
adopted star formation history in the derived stellar masses it is
worth comparing the inferred stellar mass of each galaxy for the two
adopted star formation histories, namely constant (CSF) and
exponentially decaying (EXP) star formation. We find that the agreement
 in the masses of individual objects is excellent, within the errors,
 with no obvious systematic trend or offset, indicating that the
 derived masses are robust and the adopted star formation history has
 a negligible effect on them ( $\langle$ M$_\ast$$_{CSF}$ $\rangle$ =
 10.471 $\pm$ 0.101 and $\langle$ M$_\ast$$_{EXP}$ $\rangle$ =10.451
 $\pm$ 0.126).

We note however, that the uncertainties in the mass estimates become
more serious when one assumes more complex star formation histories,
such as the superposition of a young, roughly continuous episode of
star formation and an old burst, with t$_{sf}$ $>>$ $\tau$, that
peaked sometime in the past (Papovich et al. 2001) or introducing
random bursts during the adopted star formation histories (Glazebrook
et al. 2004). These studies indicate that the use of the simple star
formation histories (CSF or EXP) is likely to provide a lower limit on
the stellar mass estimate (Shapley et al. 2005).

\section{The stellar masses of  $z$ $\sim$ 3 LBGs}

Based on the best-fit CB07 models, we derive the stellar masses of our
sample population.  The stellar masses, ages (t$_{sfr}$) and
extinction (E(B $-$ V)) that correspond to the best fit models are
listed in Table 2.  Before examining the results though, we
investigate the impact of the TP-AGB phase on the derived masses by
comparing the stellar masses obtained with CB07 and BC03 models
respectively.

\subsection{The Impact of the TP-AGB phase on the derived stellar masses}

At a redshift of $\sim$ 3 the age of the universe is $<$ 2 Gyrs,
putting an upper limit in the age of those galaxies at 1--2G yrs.  As
the AGB stars are of intermediate age ($\sim$ 1 Gyr), one would expect
to find such stars in the stellar content of $z$ $\sim$ 3 galaxies. If
this is the case, since AGB stars are the dominant bolometric and
near--IR contributors in stellar populations with ages $\sim$1 Gyr,
 the AGB--phase should be of great importance for the interpretation
 of the rest--frame near--IR Spitzer colours of these galaxies. Recent
 studies of $z$ $\sim$ 2 galaxies (e.g., Maraston et al. 2006, Eminian
 et al. 2007, Wuyts et al. 2007) have shown that incorporating the
 TP--AGB phase in the stellar synthesis population codes has resulted
 in the reduction of the estimated stellar masses and age by a factor
 of $\sim$ 2.

\begin{figure}
\centering
\includegraphics[width=8cm,height=8cm,angle=-90]{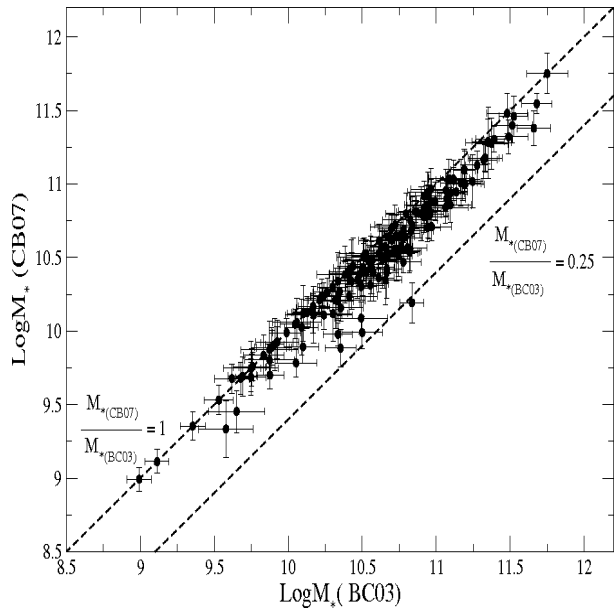}\\
\caption{\small{The derived best-fit stellar masses for the whole sample of LBGs based on CB07 over those based on BC03. 
The black dashed lines correspond to $M_{\ast(CB07)}/M_{\ast(BC03)} = 1$ and  $M_{\ast(CB07)}/M_{\ast(BC03)} = 0.25$ respectively.}}
\label{fig:mass03-07}
\end{figure}

To investigate the impact of the new AGB recipe in the derive stellar
masses of our sample, we compared the stellar masses derived from on
CB07 model to those based on BC03 models. This comparison is
illustrated in Figure \ref{fig:mass03-07} where for each LBG we plot
the stellar masses as derived from the best-fit CB07 models over those
derived based on BC03 ones.  This Figure clearly demonstrates that the
masses based on BC03 models are consistently higher than those
predicted by CB07. The mean stellar mass of the population is
$\langle$ log(M$_\ast$)$ \rangle$ = 10.621 $\pm$ 0.106 and 10.448
$\pm$ 0.099 for BC03 and CB07. This implies that the addition of the
AGB phase results in the reduction of the derived stellar masses on
average by 40\% for our whole IRAC detected sample. We note however,
that the reduction factor varies, and for some cases it can get as
high as $\sim$ 3.

\subsection{The Stellar masses}

\begin{figure} 
\centering
\includegraphics[width=8cm,height=8cm,angle=-90]{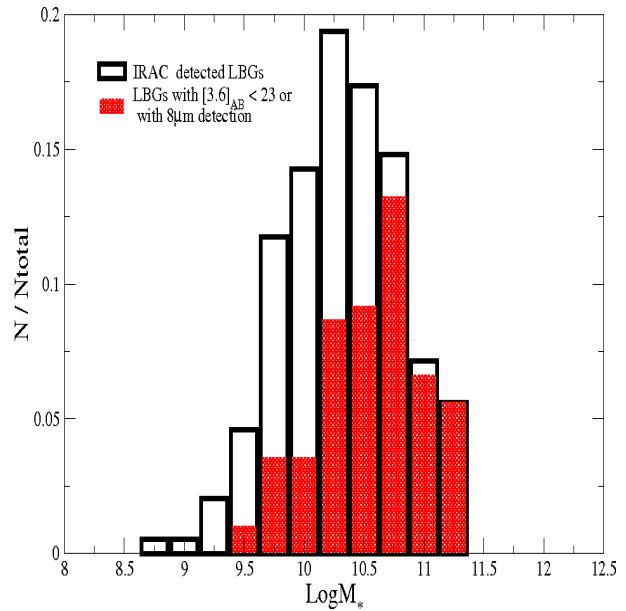}\\
\caption[Distribution of the stellar masses derived from the
best--fitting models]{\small{Distribution of the stellar masses
derived from the best--fitting models for two samples: all LBGs with
at least one IRAC detection (black) and LBGs with 8$\mu$m detection or
LBGs without 8$\mu$m detection but with $[3.6]_{AB} < 23$ (red). The
models assume a Chabrier IMF solar metallicity and the Calzetti (2000)
extinction law.}}  
\label{fig:dist} 
\end{figure}

As stated above, stellar mass is a robust parameter that is less
sensitive to uncertainties compared to other parameters. In
Magdis et al. (2008), we showed for the first time that LBGs are a
rather inhomogeneous population, ranging from those with bright IRAC
fluxes and R$-$3.6$>$1.5 to those with R$-$3.6$<$1.5 and marginal IRAC
detection. Such diversity on the rest--frame optical colours of LBGs was also presented by Shapley et al. (2005). Here we aim to translate this range in the IRAC colours
into a range of the stellar masses. As stated above, stellar mass is
a robust parameter that is less sensitive to uncertainties compared to
other parameters.  To visualize the stellar masses distribution of the
population, in Figure \ref{fig:dist} we present a histogram of the
inferred stellar masses for the whole Spitzer LBG sample and for LBGs
with 8$\mu$m detection or without 8$\mu$m detection but [3.6]$_{AB}$
$<$ 23 (i.e. extended 8$\mu$m sample).  

The majority of the massive LBGs belong to the extended 8$\mu$m sample
which has a median stellar mass of log M$_{\ast}$ = 10.711 $\pm$
0.105, while for the rest LBGs the corresponding value is log
M$_{\ast}$ = 10.169 $\pm$ 0.121. In total, there are 62 LBGs with
estimated stellar masses M$_{\ast}$ $>$ 5 $\times10$
$^{10}$M$_{\odot}$. Of these, 39 belong to the 8$\mu$m sample, 51 to
the extended 8$\mu$m sample and only 5 to the remaining LBG
sample. The same numbers for galaxies with M$_{\ast}$ $>$
10$^{11}$M$_{\odot}$ are 21, 26 and 1 respectively. Splitting the
8$\mu$m sample in two groups, those with [8.0]$_{AB}$ $>$ 22.5 and
those with [8.0]$_{AB}$ $<$ 22.5 and performing a K-S test between the
stellar masses of the two groups, reveals a significant difference
between them (P =3.14 $\times$ 10$^{-7}$). In particular, we find that
LBGs with bright 8$\mu$m fluxes are more massive, with median stellar
mass log M$_{\ast}$ = 11.017 $\pm$ 0.102, compared to log M$_{\ast}$ =
10.501 $\pm$ 0.113 for the 8$\mu$m faint LBGs.

We now focus on 24$\mu$m detected LBGs in HDFN. These LBGs are also
detected at 8$\mu$m with a median 8$\mu$m magnitude of [8.0]$_{AB}$ =
22.17 $\pm$ 0.11. The corresponding median 8$\mu$m magnitude of the
24$\mu$m undetected LBGs in HDFN is [8.0]$_{AB}$ = 23.05 $\pm$ 0.15,
indicating that they are on average fainter at 8$\mu$m than the
24$\mu$m detected LBGs.  A K-S (Kolmogorov-Smirnov) test between the
8$\mu$m magnitude distributions of the two samples confirms the
significant difference between them at a confidence level of 98$\%$ (P
=0.021). On the other hand a second K-S test between the 8$\mu$m
fluxes of the 24$\mu$m detected LBGs and the 24$\mu$m undetected LBGs
with [8.0]$_{AB}$ $<$ 22.5, reveals that the two samples are similar
at a confidence level of 68\%, showing that 24$\mu$m LBGs represent
the bright end of the 8$\mu$m sample. From the above analysis, it is
expected that 24$\mu$m LBGs are among the most massive galaxies in our
sample. Indeed, 24$\mu$m LBGs have a median stellar mass of log
M$_{\ast}$ = 10.901 $\pm$ 0.109 , while the 24$\mu$m undetected LBGs
are less massive, with median, log M$_{\ast}$ = 10.396 $\pm$ 0.101.

As discussed above, the age--dust degeneracy makes the
parameterization of these two properties uncertain and difficult to
constrain. With this in mind, it is worth noting that all LBGs with
derived ages t$_{sfr}$ $>$ 160 Myrs belong to the extended 8$\mu$m
sample.  Furthermore, the median age of the 8$\mu$m undetected, the
8$\mu$m faint ([8.0]$_{AB}$ $>$ 22.5) and 8$\mu$m bright ([8.0]$_{AB}$
$<$ 22.5) LBGs is 255.00, 404.154 and 980 Myrs respectively. Finally,
if we restrict our sample to LBGs with comparable t$_{sfr}$, we find
that those in the extended 8$\mu$m sample are consistently more
massive than those with IRAC faint colours. Since, in the same time
interval, LBGs with 8$\mu$m detection or [3.6]$_{AB}$ $<$ 23, have
grown significantly larger stellar masses, we suggest that they are
among the highest star forming LBGs or that they have undergone more
intensive star formation episodes than the rest LBGs. We note that
although the current data cannot rule out in a definitive way that the
24$\mu$m emission of these systems originates from hot dust heated by
an AGN, the lack of AGN signature in their rest-frame UV spectra as
well as their SED, that resembles cold SCUBA sources (Huang et
al. 2005), suggests that they are dominated by star formation.
 
Bringing these results together, we find that LBGs with bright IRAC
fluxes are more massive and older when compared to the rest of the
sample. This analysis has also revealed that although the addition of
the AGB--phase has reduced the estimated stellar masses of LBGs, a
substantial fraction of LBGs is still found to be massive with
M$_\ast$ $>$ 5$\times$10$^{10}$ M$_\odot$.  LBGs with 8$\mu$m
detection are among the most massive in the present sample, while at
the high end of the stellar mass distribution of the population we
find LBGs with MIPS 24$\mu$m emission.

This range in the stellar masses of the LBGs has also been
reported in previous studies (e.g Papovich et al. 2001, Shapley 2001,
Reddy et al. 2006, Rigopoulou et al. 2006). These studies however,
were based on SSP codes without a sophisticated treatment of the
AGB--phase, namely using BC03. A direct comparison with our results
indicates that these studies have overestimated the stellar masses of
the LBGs by a factor of 1.5--2. For example, Rigopoulou et al. (2006)
reports a median stellar mass for the 8$\mu$m detected LBGs of
M$_{\ast}$ = 10.98 $\pm$ 0.112, $\sim$ 1.8 times higher than that
found in our study. Hence, it becomes apparent that a fraction of the
the rest--frame NIR light that was previously attributed to a number
of relatively old stars (BC03) is now (CB07) interpreted as light
emitted from stars in the AGB phase, resulting in the reduction of the
derived stellar mass.  We conclude that although the addition of the
AGB--phase has not narrowed the range of the stellar masses of the
LBGs, it has reduced the stellar mass budget of the population and to a greater extend of all star-forming galaxies older than 0.1Gyrs. \\

\subsection{Mass to Light ratios as a function of wavelength}

It has been suggested that estimates of stellar mass from photometric
measurements become increasingly reliable as one obtains longer
wavelength data (e.g. Labbe et al. 2005), and that the addition of
longer wavelength photometric points (i.e. the IRAC bands) plays a key
role in our understanding of the properties of the LBG population as a
whole. Here, we investigate the distribution of stellar mass with
rest--frame wavelength as we move from optical to the near--infrared
bands.

\begin{figure}
\centering
\includegraphics[width=8cm,height=12cm]{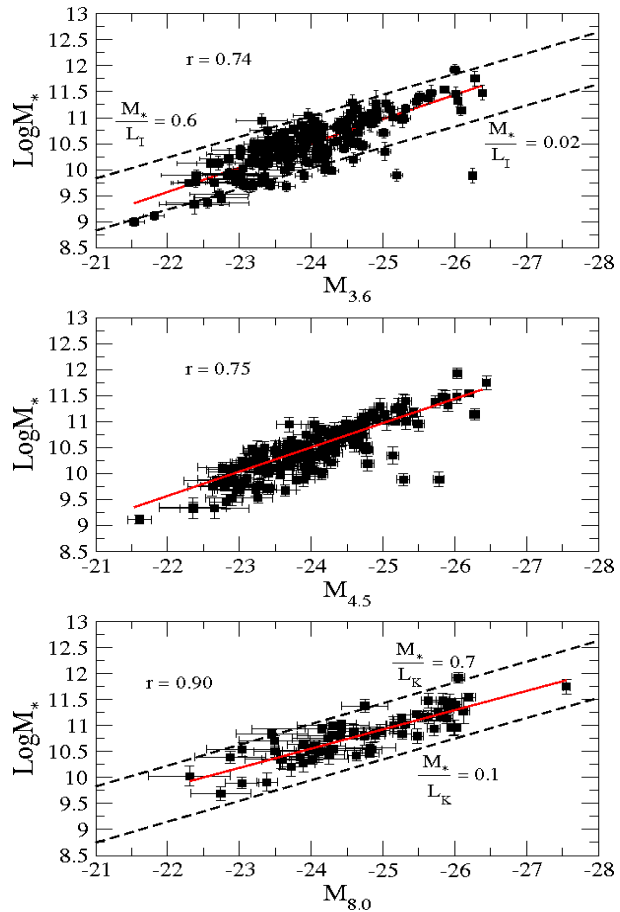}\\
\caption[Estimated stellar masses from the best fit model (for all IRAC detected LBGs) as a function of absolute 3.6$\mu$m 
magnitude and absolute 4.5$\mu$m magnitude]{\small {Stellar masses estimated from the best fit model (for all IRAC detected LBGs) as a function of absolute [3.6] $\mu$m 
magnitude, rest--frame I--band (top), absolute [4.5]$_{AB}$ magnitude (middle) and  absolute [8.0]$_{AB}$. The scatter 
in stellar mass--to--light ratio at a given 3.6$\mu$m luminosity
(rest--frame I band luminosity at redshift 3) can be as high as 30 while at 8.0$\mu$m (rest--frame K) it
is significantly reduced. The dashed lines in the top and bottom panels indicate
the range of stellar M/L in the sample and are given in solar units evaluated at rest--frame
I and K band while the red solid lines to the best linear regression fit. The r value for each regression is also noted}}
\label{fig:m-phot1}
\end{figure}

In Figure \ref{fig:m-phot1}(top) we show the distribution of stellar
masses for all IRAC detected LBGs, as a function of absolute [3.6]
$\mu$m magnitude which at the median redshift of $\sim$3 corresponds
to rest--frame 0.9$\mu$m (i.e. I--band). There is clearly a
correlation between absolute I--band magnitude and stellar mass
(correlation coefficient r=0.74) while at the bright end (where most
of the massive galaxies are found) the correlation becomes tighter
(r=0.90, if we exclude the outlier). The M/L$_{I}$ values (both M and
L normalized in solar units) range from 0.6 ($\sim$ 5 times lower than
that of the present day galaxies (Bell et al. 2003)) to 0.02,
revealing a scatter of $\sim$ 30.

The spread in the M/L$_{I}$ values can be attributed to the wide range
of star--formation histories among the LBGs and it is therefore rather
difficult to associate a single I--band rest--frame luminosity to
specific a stellar mass. The situation remains the same at 4.5$\mu$m
as shown in Figure \ref{fig:m-phot1} (middle panel).

Likewise, in Figure \ref{fig:m-phot1} bottom panel we plot the
distribution of stellar masses as a function of absolute 8.0$\mu$m
magnitude which would correspond to rest--frame $\sim$ 2.0 $\mu$m
(i.e. K--band). The correlation between stellar mass and magnitude
becomes tighter with r = 0.90 while the scatter in M/$L_{K}$ values
decreases and is now a factor of $\sim$ 7.0. The largest M/L$_{K}$
values approach that of the present day galaxies but the average is
0.23, i.e. several times smaller.  Due to the small scatter in the M/L
values of the LBGs with 8$\mu$m detection, one can crudely associate a
K--band rest--frame absolute magnitude to a specific stellar mass
through the relation : \\

Log(M$_{\ast}$/M$_{\odot}$) $=$ 2.01 ($\pm$ 0.65) $-$ 0.35($\pm$ 0.03)$\times$M$_{8.0}$.\\

\begin{figure}
\centering
\includegraphics[width=8cm,height=10cm]{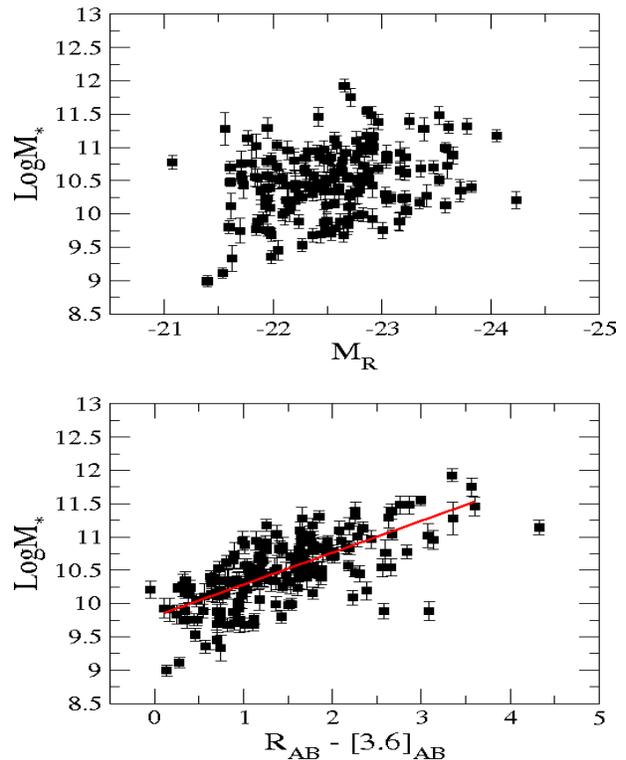}\\
\caption[Estimated stellar masses from the best fit model as a function of absolute R magnitude, and 
R $-$ 3.6$_{AB}$ colour]{\small{ Top: Stellar masses estimated from the best fit model 
as a function of absolute R magnitude. Bottom: Stellar masses estimated from the best fit model as a function 
of R $-$ [3.6]$_{AB}$ colour for LBGs in our sample. There is a strong correlation between 
masses and the R $-$ [3.6$_{AB}$] colour, particularly for the 8 $\mu$m LBG sample. 
The most massive objects (M$_\ast$ $ > 5\times 10^{10}$ M$_\odot$) tend to show the reddest R $-$ [3.6]$_{AB}$ colours as well. The red line corresponds to the best linear regression fit.}}
\label{fig:m-phot2}
\end{figure}

From these simple comparisons, one can conclude that the mid--IR bands
and especially the IRAC 8$\mu$m band (which samples the rest frame NIR
wavelengths for $z$ $\sim$ 3 LBGs) provide a more accurate estimate of
the M/L ratio compared to that obtained when using optical bands. To
further illustrate this argument, in Figure \ref{fig:m-phot2} (top
panel) we plot the distribution of stellar mass as a function of
absolute R--band magnitude, for all LBGs in the present sample with at
least one IRAC detection.  The correlation coefficient drops to 0.32
and a mass to light relation cannot be established using only optical
photometric points as the scatter in the M/L values is very large
(possibly due to different star formation histories). With the
addition of IRAC data this scatter decreases and is best correlated
with the IRAC 8 $\mu$m band. This was somewhat expected since this
band is sensitive to the light from the bulk of the stellar activity
accumulated over the galaxy's lifetime (see also Bell \& de Jong
2001).

The region of the spectrum that becomes sensitive to the ratio between
the current star formation and the integral of past star formation, is
between the UV and the NIR, conveniently measured by the observed R
$-$ [3.6]$_{AB}$ colour in the current sample. This point is
illustrated in Figure \ref{fig:m-phot2} (bottom panel), which shows
that the inferred stellar mass is well correlated with R $-$
[3.6]$_{AB}$. In fact, the correlation between the R $-$ [3.6]$_{AB}$
colour and the inferred stellar mass is almost as significant as the
correlation between stellar mass and M$_{4.5\mu m}$, with LBGs with
redder R $-$ [3.6$_{AB}$] colours having higher stellar
masses. Finally, we note that in the high end of the stellar
distribution one can find only LBGs that belong to the extended
8$\mu$m sample. It is the bright IRAC colours that disentangle between
the several star formation histories and provide a robust estimate of
the mass to light ratios. Finally, ``red'' LBGs with faint IRAC
colours and hence low stellar masses could exist and occupy the
bottom--right corner of Figure \ref{fig:m-phot2} but are not present
in the current sample due to the limiting apparent R magnitude (
$\sim$ 25.5 ) of their selection.

\section{Number  Density of Massive LBGs}

\begin{figure*} 
\centering
\includegraphics[width=10cm,height=14cm,angle=-90]{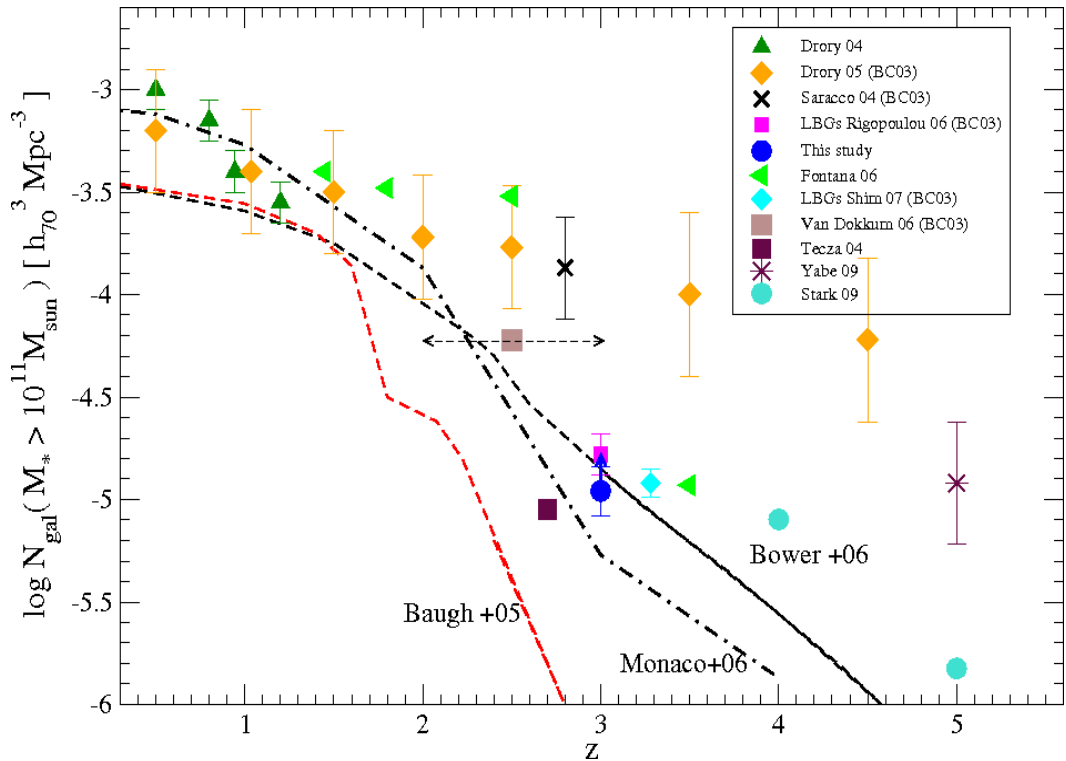}\\
\caption[Number density of massive LBGs]{\small{The number density of
galaxies with stellar masses $>$ 10$^{11}$ M$_{\odot}$ as a function
of redshift. The blue filled circle shows our result for massive LBGs,
while the blue box the result of massive LBGs by R06. Data points from
other observations are plotted with different symbols, and the
prediction from recent semi--analytic and hydrodynamic models of
galaxy formation are over-plotted (red--dashed line from Baugh et
al. 2005, black--dashed line for Bower et al. 2006,
black--dashed--dotted line for Monaco et al. 2006).}}  
\label{fig:nd}
\end{figure*}

It is interesting to find that, even after the addition of the
AGB--phase, the best--fit CB07 results indicate that $\sim$ 15\% of
the LBGs in the present sample has estimated stellar masses
(M$_{\ast}$ $>$ 10$^{11}$M$_\odot$).  These galaxies are mainly found
among the LBGs that have IRAC 8$\mu$m detection, consisting an ideal
sample of high--$z$ K-selected galaxies. In Figure \ref{fig:nd} we
plot the number density of massive (M$_{\ast}$ $>$ 10$^{11}$
M$_\odot$) LBGs in our sample as a function of redshift and compare
the value with other observational results (Drory et al. 2005; Saracco
et al. 2004; Fontana et al. 2006; Rigopoulou et al. 2006, Shim et
al. 2007, Tecza et al. 2004, Van Dokkum et al. 2006). If we assume an
effective volume for the U--dropouts V = 450 h$^{-3}$ Mpc$^{3}$
arcmin$^{-2}$ (Steidel et al. 1999), the effective co--moving volume
becomes V = 1400 Mpc$^{3}$ arcmin$^{-2}$ (the volumes have been
weighted according to the number of objects per R--magnitude bin,
$\Omega{_m}$ = 0.3, $\Omega{_\lambda}$ = 0.7 cosmology with H$_{0}$ =
70 km s$^{-1}$ Mpc$^{-1}$). For the LBGs with M$_{\ast}$ $>$ 10$^{11}$
M$_\odot$ in the 1066 arcmin$^{2}$ covered by our study, the derived
co--moving density at the average redshift $z$ $\sim$ 3, is $\Phi$ =
(1.12 $\pm$ 0.4) $\times$ 10$^{-5}$ Mpc$^{-3}$ (the uncertainty of the
derived number density is based on Poisson error). This value is lower
by a factor of $\sim$ 1.5 when compared to that found when we consider
stellar masses derived from BC03 models (1.67 $\pm$ 0.32 $\times$
10$^{-5}$ Mpc$^{-3}$). It is worth noticing that our BC03 result is in
excellent agreement with that of Rigopoulou et al. (2006) for a sample of 148 LBG in EGS.  On the other hand, for the LBGs
with M$_{\ast}$ $>$ 8$\times$10$^{10}$ M$_\odot$, we derive a
co--moving density of $\Phi$ = (1.75 $\pm$ 0.26) $\times$ 10$^{-5}$
Mpc$^{-3}$ comparable to that of the BC03 result for LBGs with
M$_{\ast}$ $>$ 10$^{11}$ M$_\odot$.  

We have to stress that due to the spectroscopic selection of the LBGs
in the present sample, we are restricted to optically bright (or
rest--frame UV-bright) LBGs (R $\leq$ 25.5).  Therefore, our sample
does not account for the optically faint (R $>$ 25.5) LBGs as well as
for other optically faint massive populations at $z$ $\sim$ 3, such as
DRGs and SMGs. Reddy \& Steidel (2009) found that the fraction of
massive galaxies among UV-faint galaxies in general is very small
(2\%) given the steep faint-end slope of the UV luminosity
function. On the other hand, Van Dokkum et al. (2006) suggests that
77\% of the massive galaxies at 2$<$$z$$<$3 are selected as DRGs and
20\% as UV-bright LBGs. Furthermore, Tecza et al. (2004), report a
number density of 8.9 $\times$ 10$^{-6}$ Mpc$^{-3}$ for massive SMGs
Given that the overlap between LBGs and the rest of the populations is
small (between DRGs and LBGs $\sim$7\% (Van Dokkum et al. (2006) while
between LBGs and SMGs is probably larger, $\sim$15\% (Chapman et
al. 2005)), we note that the above number density of massive LBGs
should be regarded as a lower limit for massive galaxies at $z$ $\sim$
3.  Despite the biases introduced by different selection techniques,
LBGs constitute the largest existing sample of star--forming galaxies
at $z$ $\sim$ 3 with confirmed spectroscopic redshift and therefore,
our result places a firm and robust lower limit on the number density
of massive galaxies at $z$ $\sim$ 3.

It is also worth comparing the present result with several theoretical
predictions from semi--analytical models.  In Figure \ref{fig:nd} it
is evident that for the case of Baugh et al. (2005) the evolution of
the number density of massive (M$_\ast$ $\sim$ 10$^{11}$ M$_\odot$)
galaxies with redshift is slower than the prediction of hierarchical
models, at least in the redshift range 0$<$$z$$<$3. On the other hand,
the discrepancy between the model and the observational constraints is
much reduced when one considers more recent predictions by Bower et
al. (2006) and Monaco et al. (2006). The galaxy formation model
presented by Bower et al. (2006) is based on the hierarchical model of
Cole et al. (2000). The key assumptions of this model are that at
high--$z$ the cooling times in halos were short enough to allow the
gas to cool on the free-fall timescale and that at the black hole
masses were lower at high--$z$.  These two assumptions make the AGN
feedback in their model less effective, allowing high--$z$ massive
galaxies to be built in the framework of the hierarchical model. We
note that the fact that Bower et al. (2006) predictions exceed the
number density of our sample is reasonable, since it allows for other
massive galaxies, that are not selected with the Lyman Break
technique, to exist at $z$$\sim$3.

\subsection{Stellar Mass Density of  $z$ $\sim$ 3 LBGs}

Determining the evolution of the global stellar mass with redshift is
one of the most challenging tasks of modern cosmology. Hence, it would
be interesting to estimate the Stellar Mass Density (SMD) of our
sample and attempt to place constraints on this property. Given the
depth of our observations, our sample is complete at stellar masses
M$_\ast$ $>$ 2 $\times$ 10$^{10}$M$_\odot$.  Therefore, we choose to integrate
from log(M$_\ast$/M$_\odot$ )=10.301 to log(M$_\ast$/M$_\odot$)= 13
and derive a SMD $\rho$ = 7.08 $\pm$0.8 $\times$ 10$^{6}$
M$_\odot$Mpc$^{-3}$. A comparison with the SMD at $z$ $\sim$ 0.1 as
estimated by Cole et al. 2001, indicates that UV-bright LBGs with
M$_\ast$ $>$ 2 $\times$ 10$^{10}$M$_\odot$ at $z$ $\sim$3 harbor
$\sim$2.5\% of the total stellar mass seen in the local
universe. However, we have to stress that our sample is censored both
at the high and low mass end: As discussed above, the Lyman Break
selection technique misses $\sim$80\% of the massive galaxies at the
redshift range $z$=2-3, although the actual fraction at $z$ $=$ 3 has
not yet been quantified.  On the other hand, focusing on the LBGs with
IRAC detection introduces a bias against LBGs of lower stellar mass.
Furthermore, as pointed out by Reddy $\&$ Steidel (2009), apart from
the luminous and massive high--$z$ galaxies, the faint--UV and less
massive galaxies could also play an important role in the stellar mass
assembly at these redshifts. More particularly, they claim that, for
galaxies with M$_{\ast}$ $<$ 10$^{11}$M$_\odot$, the total stellar
mass contained in UV--faint is roughly equal to that contained in
UV-bright galaxies.  Hence, the above value is a lower limit not only
for the whole $z$ $\sim$ 3 galaxies but also for the
U-dropouts. Although we cannot make any safe assumptions about the
fraction of massive LBGs and thus correct for the stellar masses
residing in massive galaxies at $z$ $\sim$ 3, if we account for the
contribution of the UV-faint galaxies to the stellar mass assembly, we
find a SMD $\rho$ $\sim$ 1.2 $\times$ 10$^{7}$ M$_\odot$Mpc$^{-3}$ for
the UV selected galaxies at $z$ $\sim$ 3.

\begin{figure} 
\centering
\includegraphics[width=8cm,height=8cm,angle=-90]{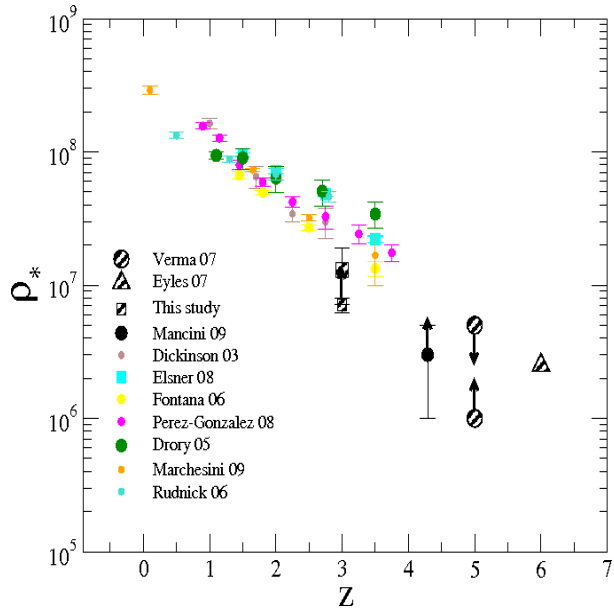}\\
\caption[Number density of massive LBGs]{\small{Evolution of the
stellar mass density with redshift. The y axis indicates the stellar
mass density of galaxies at various redshifts, as estimated from
several studies. Shadow-black symbols correspond to studies focused on
LBGs. When necessary, values have been converted to a Chabrier
IMF. For our study, we show the lower limit for UV-bright LBGs as well
as the estimated $\rho$ for the whole (bright$+$ faint) UV--selected
population of $z$ $\sim$3 galaxies (shadow-black boxes).}}
\label{fig:r} 
\end{figure}

We now compare the SMD of our sample with that of other contemporary
studies.  In Figure \ref{fig:r}, we plot the evolution of the total
stellar mass density as a function of redshift for several samples
drawn from the literature (Fontana et al. 2006, Dickinson et al. 2003,
Rudnick et al. 2006, Drory et al. 2005, Perez-Gonzalez et al. 2008,
Elsner et al. 2008, Marchesini et al. 2009 (MUSYC), Verma et al. 2007,
Eyles et al. (2007), and Mancini et al. 2009).  The values from the
literature derived assuming a Salpeter (1955) IMF have been scaled to
a pseudo-Chabrier IMF by dividing the stellar mass densities by a
factor of 1.8. Although the values from the literature are either at
lower or higher $z$ than our sample, we note that our value is broadly
consistent with previous studies.  The fact that, even after
correcting for the contribution of faint-UV galaxies, our value is
lower by a factor of $\sim$ 2 when compared to the general trend can
be attributed to the stellar masses of optically faint massive
galaxies (i.e DRG,SMGs) and to systematics introduced by the different
methods used to derive the stellar masses, especially the treatment of
the AGB phase.

It is also interesting to compare our result with that of Verma et
al. (2007), since the photometric criteria for selecting LBGs at $z$
$\sim$ 5 in their study were chosen to match that of Steidel et
al. (2003) for our sample. It seems that LBGs at $z$ $\sim$3 have
assembled 2--4 times more stellar mass than their high--$z$
counterparts. We note again, that this type of analysis suffers from
large uncertainties introduced by selection biases, cosmic variance
and different methods of stellar mass determination, making it
difficult to securely draw conclusions.  On the other hand, LBGs at
$z$ $\sim$3 consist of the largest existing sample of high--$z$
galaxies with confirmed spectroscopic redshift and hence their study
provides robust constraints on the nature of high--$z$ galaxies.

\section{The star formation mass correlation at $z$ = 3}

Recently, Daddi et al. (2008), have shown that the UV-corrected star
formation rate and stellar mass define a tight correlation in
K-selected galaxies at $z$ $\sim$ 2.  Similar results have been
reached by Noeske et al. (2007) and Elbaz et al. (2007) for galaxies
at $z$ $\sim$ 1 and at $z$ = 0 based on data drawn from the Sloan
Digital Sky Survey (Elbaz et al. 2007), although with a lower
normalization reflecting the overall decline in cosmic SFR density
with time (Madau et al. 1996). On the other hand, Daddi et al. (2009),
based on an IRAC sample of B--dropouts, showed that the locus of
typical massive $z$ $\sim$ 4 B-dropouts doesn't support a continuously
increasing specific star formation rate (SSFR defined as the current
SFR per unit stellar mass, $\phi$) with redshift, suggesting instead a
plateau of the SSFR at $z$ $>$ 2. Here, we try to fill the gap between
$z$=2 and $z$=4 and examine the SFR-stellar mass relation at $z$=3.

To determine the UV-corrected star-formation rate we use the UV
luminosity, assuming a CSF, Calzetti (2000) extinction law and solar
metallicity. At $z$ $\sim$ 3 the observed R and G bands correspond to
rest--frame 1600$\AA$ and 1200$\AA$ respectively. We use these
measurements to derive the $\beta$ slope of the UV spectrum for each
galaxy, after applying a correction for the IGM attenuation that
follows the prescription of Madau et al. (1995). Then, based on the
best-fit CB07 model SED derived as discussed in section 3.1, we apply
a K-correction to derived the observed flux at 1500$\AA$ and a
Calzetti (2000) law to estimate the intrinsic L$_{1500}$ for each
galaxy in our sample. To convert the intrinsic L$_{1500}$ to
star-formation rate we use the relation adopted by CB07 models :

\begin{equation}{
{\rm SFR} (M_\odot{\rm yr}^{-1}) = L_{1500 \AA} [{\rm erg\ s^{-1} Hz^{-1}}] / (8.85\times10^{27})}
\end{equation}
\begin{figure}
\centering
\subfigure{
\includegraphics[width=7cm,height=8cm,angle=-90]{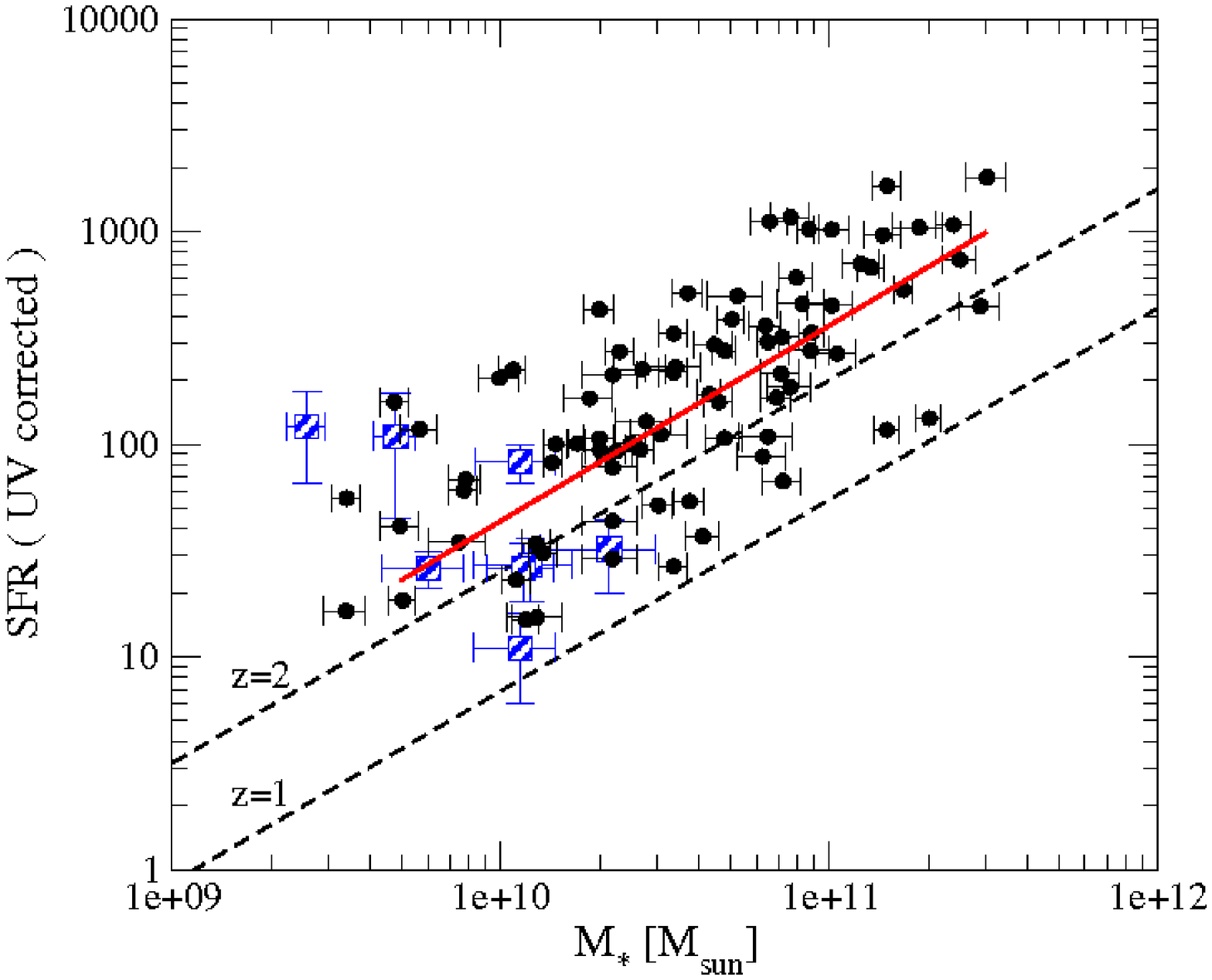}}\\
\subfigure{
\includegraphics[width=4cm,height=7.5cm,angle=-90]{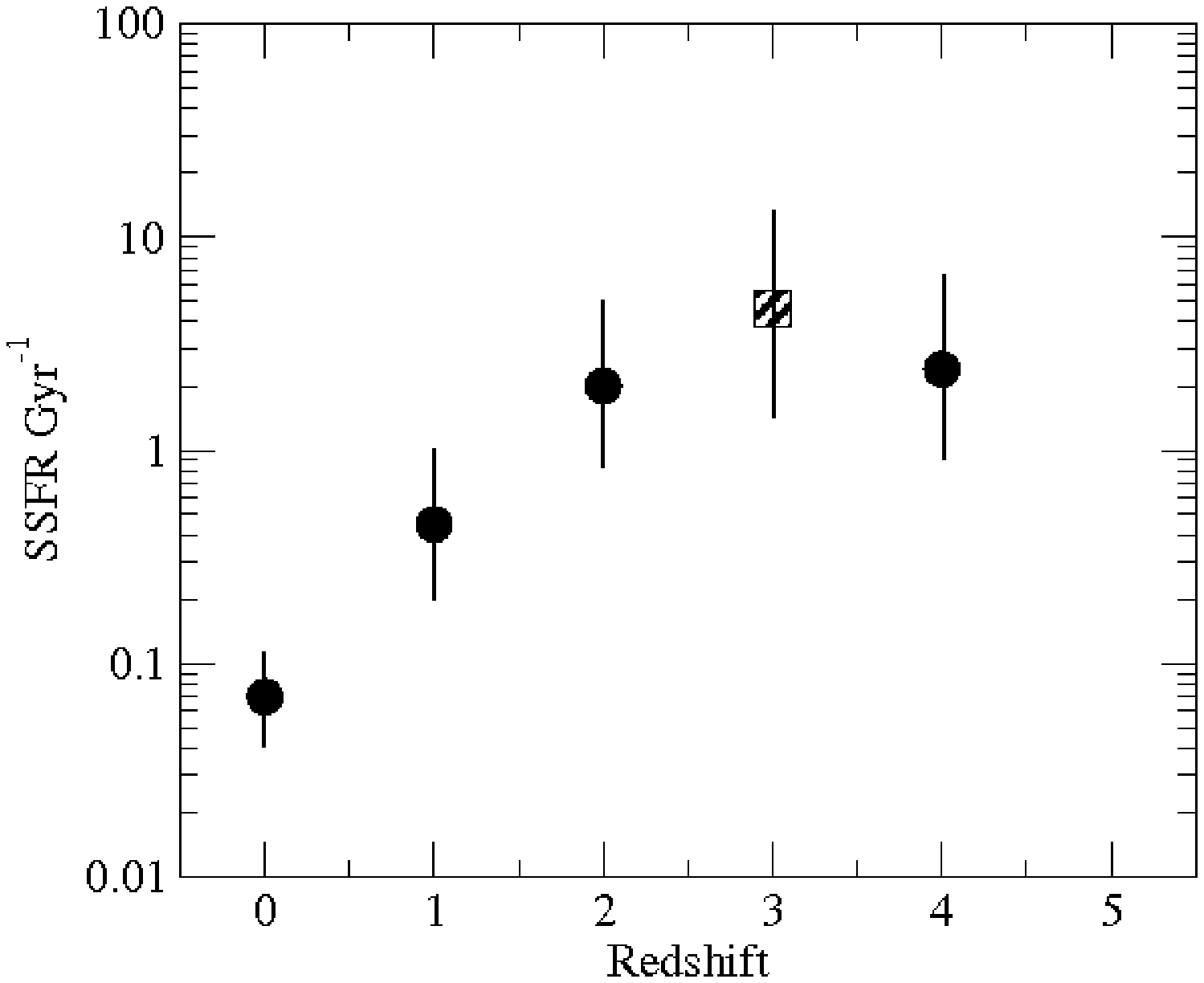}}\\

\caption{\small{Upper: The UV-corrected SFR over the stellar masses for our sample. The red line corresponds to the linear regression best-fit (r=0.70) with a slope $\alpha$ =0.91. The black-dashed lines correspond to the SFR-M$_{\star}$ relation relation at $z$=2 and $z$=1 by Daddi et al. (2008) and Elbaz et al. (2007) respectively. Blue boxes correspond to H$\beta$ based SFRs for LBGs at $z$ $\sim$3 by Mannucci et al. 2009. Bottom: Specific star formation rate (SSFR = SFR/M$_\star$) as a function of redshift. Black circles show the average SSFR at a stellar mass of 5$\times$10$^{10}$ M$_{\odot}$ from Daddi et al. (2007),(2009) and Elbaz et al. (2007), while the black--shadowed box corresponds to the measurement of the current study for galaxies at $z$=3. The vertical lines show the measured 1$\sigma$ range.}}
\label{fig:sfr}
\end{figure}
   
Before going any further we have to stress that this technique suffers
from several limitations.  In applying a reddening correction the
technique assumes that the UV spectral slope is entirely due to
reddening, rather than to the presence of evolved stellar populations.
Furthermore, the strongest starbursts are opaque to UV radiation and
their total SFR activity cannot thus be reliably estimated solely from
the UV, even after reddening corrections (e.g., Goldader et al. 2002;
Buat et al. 2005, Reddy et al. 2006).  On the other hand, Rigopoulou
et al. (2006) argued that the fact that a fraction of LBGs is detected
at 24$\mu$m implies the existence of significant amounts of dust. This
indicates that dust and not evolved stellar populations is responsible
for the UV spectral slope, favoring a CSF history. Moreover, Carilli
et al. (2008), based on stacking radio analysis of U-dropouts from the
Cosmos field, found that the SFR implied by the radio luminosities are
larger by a factor of $\sim$ 1.8 when compared to that derived from UV
luminosities without correcting for dust extinction.
 
Finally, Chapman et al. (2009) and Rigopoulou et al. (2009) (in
preparation) using SCUBA and IRAM observations both report on the
detection of LBGs in sub-mm. In particular, Chapman et al. (2009)
finds a good agreement between UV and S$_{850}$ derived SFR at faint
sub-mm levels and suggest that LBGs may contribute significantly to
the source counts of sub-mm selected galaxies in the 1-2mJy
regime. This indicates that considerable amounts of dust are found in
LBGs. Evidence, that LBGs contain significant amounts of dust have
also been provided by several other studies such as Adelberger \&
Steidel (2001), Papovich et al. (2001), Shapley et al. (2001), Reddy
et al. (2006, 2008).  This, along with the fact that LBGs are (from
their selection) blue actively star-forming galaxies, enforces the
assumption of a UV spectrum reddened by the presence of dust rather
than by a decaying SFR.  Hence, the main source of uncertainty in our
approach should be the adopted extinction law and the geometry of the
distribution of the dust.

Having discussed the uncertainties, in Figure \ref{fig:sfr} we show
the SFR--stellar mass relation using the UV corrected SFRs and the
stellar masses as estimated by our CB07 analysis assuming a CSF
history. We restrict our sample to LBGs at 2.8$<$$z$$<$3.2, detected
at both 3.6 and 4.5 $\mu$m and $\Delta$R $<$0.2 ( where $\Delta$R is
the error of the R$_{AB}$ magnitude) to reduce the uncertainty on the
estimated $\beta$ slope, and therefore the SFR.  A linear regression
fit to our data suggests a relatively tight correlation (r = 0.70)
with a logarithmic slope $\sim$0.91 (although with considerable
scatter), implying that more massive LBGs tend to have higher
star-formation rates.  The slope of this correlations is similar to
that at lower redshifts (Daddi et al. 2008, Elbaz et al. 2007, Noeske
et al. 2007) but with higher normalization factor. On the other hand,
the current normalization factor is higher than that of the B-dropouts
at $z$ $\sim$ 4. With a median specific star formation rate, of
$\sim$4.5Gyrs$^{-1}$, which is larger by a factor of two than that of
$z$ $\sim$2 and $z$ $\sim$4 samples, it seems that the evolution of
the SSFR peaks at $z$ $\sim$ 3 and then drops towards lower redshifts
(Figure \ref{fig:sfr} bottom).

Contrary to our findings, Mannucci et al. (2009), based on SFRs
derived from H$\beta$ fluxes of LBGs, argued that $z$ $\sim$3 LBGs do
not show a correlation between SFR and stellar mass. To investigate
this discrepancy we include their results in Figure
\ref{fig:sfr}(top). We see that the small number as well as the small
range of the stellar masses of the LBGs in their sample are not
sufficient to reveal any correlation. As a matter of fact their result
is in agreement with ours for a small stellar mass bin. Indeed, if we
restrict our sample to LBGs with 9$<$logM$_\ast$/M$_\odot$$<$10.3, the
scatter in their sample is similar to the one we find, indicating that
there is no correlation between SFR and stellar mass.  It is the large
numbers and the large stellar mass range of our sample that enables
the detection of the correlation. A similar trend although with
shallower slope is also reported by Erb et al. (2006) for a sample of
$z$ $\sim$2 UV selected galaxies.  We conclude that although further
investigation is required, the SFR-stellar mass relation seen in lower
redshifts, seems to hold for $z$ $\sim$ 3 LBGs.  Finally, one should
note that the average SFR found for our sample ($>$
100M$_{\odot}$yr$^{1}$), should be regarded as typical only for IRAC
detected and hence more massive LBGs.

\section{Summary}

Using IRAC photometry for a robust sample of 196 LBGs at $z$ $\sim$3
with confirmed spectroscopic redshift, we carried out a detailed
mid-IR study of the LBG population.  For our analysis we used both the
new CB07 code that incorporates an updated prescription of the AGB
phase as well as the widely used BC03. The following results were
reached:

$\bullet$ {\bf{Stellar masses}} Based on results derived by the CB07
code, we constrained the properties of the population. We find
that although the addition of the AGB phase has resulted in the
reduction of the derived stellar masses, on average by a factor of
$\sim$ 1.4, the range of the stellar masses of the LBGs is similar to
that found by previous studies. In particular, we find that the
stellar masses of the population span from M$_\ast$ $\sim$ 10$^{9}$
M$_\odot$ for LBGs with faint IRAC colours to M$_\ast$ $\sim$
10$^{11}$ M$_\odot$ for a fraction of LBGs detected at 8$\mu$m and for
ILLBGs.  The inferred ages of these massive systems are considerably
higher than the rest of the population, indicating that they have been
star-forming for $\sim$ 1 Gyr. We show how the stellar mass correlates
with wavelength and find that IRAC bands improve dramatically the
accuracy of the derived M/L ratios when compared to that obtained when
using optical bands. Finally, we show that LBGs with redder
R-[3.6]$_{AB}$ colours have higher stellar masses.

$\bullet$ {\bf{Number Density and Stellar mass Density of massive
LBGs}} We find that even after the addition of the AGB phase in the
SED models, a considerable fraction of LBGs ($\sim$15\%) is massive,
with stellar masses M$_\ast$ $>$ 10$^{11}$M$_\odot$. We calculate the
number density of these LBGs and find $\Phi$= (1.12 $\pm$ 0.4)
$\times$ 10$^{-5}$ Mpc$^{-3}$. This is $\sim$1.5 times lower than that
predicted from previous studies, providing a better match with current
theoretical models. The stellar mass density of IRAC detected,
optically bright LBGs is $\rho$ = 7.08 $\pm$0.8 $\times$ 10$^{6}$
M$_\odot$Mpc$^{-3}$, indicating a lower limit for the stellar mass
density of the whole UV selected population of $z$=3 galaxies, $\rho$
$\sim$ 1.2$\times$ 10$^{5}$ M$_\odot$Mpc$^{-3}$. Comparing our result
with values from the literature, we find that LBGs at $z$=3 have
assembled 2-4 times more stellar mass than their high-$z$ siblings.

$\bullet$ {\bf{Star-formation--Mass correlation at $z=3$}} We find a
relatively tight correlation between the UV--corrected SFR and stellar
mass for galaxies at $z$=3. This correlation has a similar slope to
that found at other redshifts, but a higher normalization factor. We
find an average SSFR of 4.6Gyr$^{-1}$ for our sample, indicating that
the evolution of the SSFR peaks at $z$=3 and drops at lower redshifts.

\section{Acknowledgments} This work is based on observations made with
the Spitzer Space Telescope, which is operated by the Jet Propulsion
Laboratory, California Institute of Technology under a contract with
NASA. Support for this work was provided by NASA through an award
issued by JPL/Caltech. GEM would like to thank S. Charlot for
providing the new CB07 code as well as D. Elbaz, H. Aussel and
E. Daddi for useful discussions.

\clearpage

\begin{deluxetable}{lccccccccc}

\tabletypesize{\normalsize}
\tablewidth{0pc}
\tablecaption{MIR Photometric catalogue (AB magnitude) of LBGs with confirmed spectroscopic $z$ }
\tablehead{

\colhead{Name} &
\colhead{$z$} &
\colhead{U$_{n}$-G} &
\colhead{G-R} &
\colhead{R} &
\colhead{3.6$\mu$m} &
\colhead{4.5$\mu$m} &
\colhead{5.8$\mu$m} &
\colhead{8.0$\mu$m} }

\startdata
B20902-C10          &   2.752  &   2.38  &   0.70  &  25.06  &  23.19  $\pm$   0.19  &  23.20  $\pm$   0.16  &   -  &  24.15  $\pm$   0.50 \\
B20902-C11          &   3.352  &   2.32  &   0.77  &  24.69  &  24.01  $\pm$   0.41  &   -  &   -  &   - \\
B20902-C5           &   3.098  &   2.47  &   0.93  &  24.70  &  23.71  $\pm$   0.36  &  22.99  $\pm$   0.08  &   -  &  21.32  $\pm$   0.09 \\
B20902-C6           &   3.099  &   3.45  &   0.45  &  24.13  &   -  &   -  &   -  &   - \\
B20902-C7           &   3.195  &   3.16  &   0.37  &  24.52  &   -  &  24.13  $\pm$   0.37  &   -  &   - \\
B20902-C8           &   2.970  &   2.90  &   0.76  &  24.40  &   -  &   -  &   -  &   - \\
B20902-C9           &   3.354  &   3.00  &   1.03  &  24.39  &  23.18  $\pm$   0.19  &  22.93  $\pm$   0.08  &   -  &   - \\
B20902-D11          &   2.835  &   1.81  &   0.29  &  22.97  &  21.71  $\pm$   0.11  &  21.59  $\pm$   0.09  &   -  &  21.20  $\pm$   0.08 \\
B20902-D14          &   2.766  &   2.20  &   0.59  &  24.40  &  22.38  $\pm$   0.10  &  22.36  $\pm$   0.11  &   -  &  23.53  $\pm$   0.50 \\
B20902-D8           &   2.867  &   1.80  &   0.24  &  24.24  &  23.52  $\pm$   0.20  &  22.81  $\pm$   0.08  &   -  &   - \\
B20902-D9           &   3.024  &   2.38  &   0.20  &  25.21  &   -  &   -  &   -  &   - \\
B20902-M11          &   3.303  &   2.65  &   1.18  &  24.19  &  23.00  $\pm$   0.18  &  22.74  $\pm$   0.10  &   -  &   - \\
B20902-M8           &   3.205  &   2.02  &   0.71  &  25.48  &   -  &   -  &   -  &   - \\
B20902-MD16         &   2.732  &   1.74  &   0.60  &  24.34  &  22.66  $\pm$   0.09  &   -  &   -  &  22.71  $\pm$   0.50 \\
B20902-MD21         &   2.986  &   2.16  &   1.06  &  24.18  &  22.23  $\pm$   0.10  &  22.49  $\pm$   0.11  &   -  &  22.87  $\pm$   0.98 \\
B20902-MD24         &   2.904  &   1.25  &   0.25  &  25.20  &   -  &   -  &   -  &   - \\
B20902-MD25         &   2.893  &   2.07  &   0.62  &  23.81  &  22.62  $\pm$   0.09  &  22.55  $\pm$   0.10  &   -  &   - \\
B20902-MD28         &   2.917  &   2.18  &   0.87  &  24.62  &  22.52  $\pm$   0.09  &  22.29  $\pm$   0.11  &   -  &  22.61  $\pm$   0.50 \\
C10-Q1700           &   2.919  &   2.93  &   0.90  &  24.59  &   -  &   -  &   -  &   - \\
C23-Q1700           &   3.256  &   3.31  &   0.70  &  23.99  &  23.13  $\pm$   0.26  &  23.06  $\pm$   0.15  &  23.19  $\pm$   0.40  &   - \\
C26-Q1700           &   2.904  &   2.88  &   0.70  &  24.46  &  22.88  $\pm$   0.09  &  22.74  $\pm$   0.11  &  22.49  $\pm$   0.21  &  23.13  $\pm$   0.22 \\
C7-Q1700            &   3.030  &   2.75  &   0.95  &  24.83  &  23.05  $\pm$   0.18  &  22.70  $\pm$   0.11  &   -  &  22.61  $\pm$   0.21 \\
C9-Q1700            &   2.929  &   2.93  &   0.62  &  24.56  &  22.67  $\pm$   0.10  &  22.43  $\pm$   0.10  &  23.08  $\pm$   0.64  &   - \\
D17-Q1700           &   3.127  &   1.65  &   0.10  &  24.99  &  22.32  $\pm$   0.13  &  22.11  $\pm$   0.10  &  22.18  $\pm$   0.34  &  22.60  $\pm$   0.21 \\
D19-Q1700           &   2.845  &   2.11  &   0.33  &  25.06  &   -  &   -  &   -  &   - \\
D20-Q1700           &   3.010  &   2.69  &   1.04  &  24.10  &  22.45  $\pm$   0.13  &  22.26  $\pm$   0.10  &  22.30  $\pm$   0.21  &  21.87  $\pm$   0.10 \\
D23-Q1700           &   2.861  &   2.20  &   0.68  &  24.11  &  22.88  $\pm$   0.09  &  22.79  $\pm$   0.11  &  22.98  $\pm$   0.64  &  22.77  $\pm$   0.21 \\
D25-Q1700           &   2.905  &   1.91  &   0.23  &  23.81  &  23.00  $\pm$   0.09  &   -  &  22.83  $\pm$   0.64  &  22.99  $\pm$   0.20 \\
DSF2237b-D19        &   3.265  &   2.14  &   0.19  &  25.08  &  23.08  $\pm$   0.15  &  22.90  $\pm$   0.10  &   -  &   - \\
DSF2237b-D28        &   2.932  &   1.91  &   0.32  &  24.46  &  23.64  $\pm$   0.20  &  23.34  $\pm$   0.19  &   -  &   - \\
DSF2237b-M31        &   3.392  &   2.12  &   0.90  &  24.98  &   -  &   -  &   -  &   - \\
DSF2237b-MD72       &   2.399  &   1.13  &   0.13  &  24.31  &  22.81  $\pm$   0.08  &  22.82  $\pm$   0.10  &   -  &   - \\
DSF2237b-MD81       &   2.823  &   1.32  &   0.31  &  24.16  &  21.17  $\pm$   0.06  &  20.82  $\pm$   0.07  &  20.73  $\pm$   0.09  &  20.83  $\pm$   0.10 \\
HDF-C11             &   3.218  &   2.50  &   0.86  &  24.41  &   -  &   -  &   -  &   - \\
HDF-C14             &   2.981  &   2.20  &   0.49  &  25.18  &  22.11  $\pm$   0.09  &  21.98  $\pm$   0.10  &  21.89  $\pm$   0.09  &  21.71  $\pm$   0.13 \\
HDF-C17             &   3.163  &   1.92  &   0.29  &  25.42  &  23.75  $\pm$   0.14  &  23.84  $\pm$   0.13  &   -  &   - \\
HDF-C18             &   3.148  &   2.37  &   0.31  &  25.16  &   -  &   -  &   -  &   - \\
HDF-C22             &   3.126  &   2.96  &   0.66  &  23.58  &   -  &  23.62  $\pm$   0.12  &   -  &   - \\
HDF-C24             &   3.328  &   2.39  &   0.76  &  24.33  &  24.16  $\pm$   0.38  &  23.91  $\pm$   0.09  &  24.39  $\pm$   2.00  &  23.64  $\pm$   0.15 \\
HDF-C25             &   2.973  &   2.18  &   0.39  &  24.88  &  23.84  $\pm$   0.14  &  23.89  $\pm$   0.13  &  23.70  $\pm$   0.26  &  23.30  $\pm$   0.16 \\
HDF-C26             &   3.239  &   2.70  &   1.09  &  24.25  &  23.00  $\pm$   0.09  &  22.30  $\pm$   0.08  &  22.06  $\pm$   0.14  &  22.06  $\pm$   0.14 \\
HDF-C27             &   2.940  &   2.42  &   0.60  &  24.57  &  22.95  $\pm$   0.09  &  22.83  $\pm$   0.13  &  22.92  $\pm$   0.12  &  22.39  $\pm$   0.13 \\
HDF-C28             &   3.130  &   2.68  &   1.09  &  23.50  &  20.64  $\pm$   0.05  &  21.00  $\pm$   0.04  &  21.38  $\pm$   0.09  &  21.39  $\pm$   0.11 \\
HDF-C5              &   2.664  &   2.23  &   0.42  &  24.88  &  23.10  $\pm$   0.13  &  23.32  $\pm$   0.16  &   -  &   - \\
HDF-C6              &   3.451  &   2.74  &   0.55  &  24.34  &  24.09  $\pm$   0.38  &  23.98  $\pm$   0.09  &   -  &   - \\
HDF-C7              &   2.658  &   2.46  &   0.62  &  24.57  &   -  &   -  &   -  &   - \\
HDF-C8              &   2.988  &   2.35  &   0.70  &  24.38  &  23.37  $\pm$   0.12  &  23.39  $\pm$   0.16  &  23.53  $\pm$   0.26  &  23.35  $\pm$   0.16 \\
HDF-D10             &   2.970  &   1.76  &   0.04  &  25.39  &   -  &   -  &   -  &   - \\
HDF-D11             &   2.930  &   1.42  &  -0.09  &  25.33  &  24.37  $\pm$   0.57  &  24.40  $\pm$   0.34  &   -  &   - \\
HDF-D12             &   2.856  &   1.83  &   0.32  &  24.84  &   -  &   -  &   -  &   - \\
HDF-D13             &   3.087  &   2.49  &   0.70  &  23.98  &  22.19  $\pm$   0.09  &  22.37  $\pm$   0.08  &  22.47  $\pm$   0.14  &  22.43  $\pm$   0.13 \\
HDF-D14             &   2.962  &   2.32  &   0.07  &  25.09  &  24.32  $\pm$   0.57  &  24.07  $\pm$   0.23  &   -  &   - \\
HDF-D15             &   3.131  &   2.65  &   0.62  &  23.61  &  23.29  $\pm$   0.12  &  23.27  $\pm$   0.16  &  23.22  $\pm$   0.17  &  23.13  $\pm$   0.16 \\
HDF-D2              &   2.806  &   2.07  &   0.18  &  24.49  &  23.59  $\pm$   0.11  &  23.62  $\pm$   0.12  &   -  &   - \\
HDF-D3              &   2.943  &   2.18  &   0.58  &  24.25  &  22.90  $\pm$   0.09  &  22.87  $\pm$   0.13  &  22.93  $\pm$   0.12  &  22.92  $\pm$   0.15 \\
HDF-D6              &   2.925  &   2.00  &   0.14  &  25.40  &  24.65  $\pm$   0.49  &  24.67  $\pm$   0.47  &   -  &   - \\
HDF-D7              &   2.394  &   2.11  &   0.34  &  24.55  &  22.30  $\pm$   0.10  &  22.25  $\pm$   0.08  &  22.34  $\pm$   0.14  &  22.21  $\pm$   0.13 \\
HDF-D8              &   2.410  &   2.21  &   0.26  &  25.04  &  23.93  $\pm$   0.14  &  24.02  $\pm$   0.32  &   -  &   - \\
HDF-M16             &   2.939  &   2.26  &   0.77  &  24.55  &  23.71  $\pm$   0.14  &  23.29  $\pm$   0.16  &  23.14  $\pm$   0.17  &  23.42  $\pm$   0.29 \\
HDF-M17             &   2.932  &   2.03  &   1.00  &  24.46  &  22.11  $\pm$   0.09  &  22.00  $\pm$   0.10  &   -  &  21.77  $\pm$   0.13 \\
HDF-M18             &   2.929  &   2.37  &   1.00  &  24.10  &  21.91  $\pm$   0.09  &  21.76  $\pm$   0.07  &  21.68  $\pm$   0.09  &  21.39  $\pm$   0.11 \\
HDF-M21             &   2.926  &   2.16  &   0.84  &  24.49  &  22.59  $\pm$   0.10  &  22.43  $\pm$   0.08  &  22.45  $\pm$   0.14  &  22.12  $\pm$   0.14 \\
HDF-M22             &   3.196  &   1.89  &   0.78  &  25.08  &  23.86  $\pm$   0.14  &  23.80  $\pm$   0.13  &   -  &   - \\
HDF-M23             &   3.214  &   2.43  &   1.09  &  24.61  &  21.01  $\pm$   0.09  &  21.16  $\pm$   0.07  &  21.08  $\pm$   0.07  &  21.10  $\pm$   0.07 \\
HDF-M25             &   3.106  &   2.12  &   0.74  &  24.82  &  23.61  $\pm$   0.11  &   -  &   -  &   - \\
HDF-M27             &   3.242  &   2.19  &   0.94  &  24.53  &  23.35  $\pm$   0.12  &  23.35  $\pm$   0.16  &   -  &  23.05  $\pm$   0.22 \\
HDF-M28             &   3.371  &   1.78  &   0.70  &  25.04  &  24.47  $\pm$   0.57  &  24.67  $\pm$   0.47  &  25.31  $\pm$   0.50  &   - \\
HDF-M32             &   3.363  &   1.99  &   0.76  &  24.95  &  23.83  $\pm$   0.14  &  23.67  $\pm$   0.12  &  23.91  $\pm$   0.48  &  23.53  $\pm$   0.29 \\
HDF-M35             &   3.229  &   2.20  &   1.17  &  23.98  &  22.57  $\pm$   0.10  &  22.43  $\pm$   0.08  &  22.28  $\pm$   0.14  &  22.14  $\pm$   0.14 \\
HDF-M7              &   2.990  &   2.13  &   0.67  &  24.79  &  23.63  $\pm$   0.11  &  23.79  $\pm$   0.13  &   -  &   - \\
HDF-M9              &   2.975  &   2.15  &   0.78  &  24.73  &  22.55  $\pm$   0.10  &  22.36  $\pm$   0.08  &  22.33  $\pm$   0.14  &  21.76  $\pm$   0.13 \\
HDF-MD10            &   2.979  &   1.43  &   0.15  &  25.17  &  24.62  $\pm$   0.49  &  24.09  $\pm$   0.23  &   -  &   - \\
HDF-MD18            &   2.442  &   1.17  &   0.03  &  24.98  &  24.28  $\pm$   0.38  &  24.21  $\pm$   0.23  &   -  &   - \\
HDF-MD19            &   2.931  &   1.84  &   0.63  &  24.66  &  23.40  $\pm$   0.11  &  23.38  $\pm$   0.16  &  23.44  $\pm$   0.26  &  23.17  $\pm$   0.16 \\
HDF-MD22            &   3.194  &   1.79  &   0.67  &  24.53  &  24.19  $\pm$   0.38  &  24.13  $\pm$   0.23  &  24.23  $\pm$   1.32  &  23.99  $\pm$   0.24 \\
HDF-MD37            &   2.830  &   1.48  &   0.24  &  24.92  &  23.97  $\pm$   0.14  &  24.14  $\pm$   0.32  &   -  &  24.72  $\pm$   0.57 \\
HDF-MD3             &   2.898  &   1.82  &   0.71  &  23.86  &  20.77  $\pm$   0.06  &  21.25  $\pm$   0.07  &  21.56  $\pm$   0.09  &  21.94  $\pm$   0.10 \\
HDF-MD40            &   2.482  &   2.00  &   0.65  &  24.94  &  22.38  $\pm$   0.10  &  22.25  $\pm$   0.08  &  22.11  $\pm$   0.14  &  22.17  $\pm$   0.14 \\
HDF-MD45            &   2.345  &   1.83  &   0.69  &  23.55  &  21.04  $\pm$   0.09  &  22.78  $\pm$   0.11  &  21.32  $\pm$   0.09  &  21.19  $\pm$   0.07 \\
HDF-oC14            &   2.928  &   1.21  &   0.36  &  25.61  &   -  &   -  &   -  &   - \\
HDF-oC26            &   3.182  &   1.66  &   0.40  &  25.63  &  25.49  $\pm$   0.14  &   -  &   -  &   - \\
HDF-oC29            &   3.161  &   1.59  &   0.64  &  25.49  &  25.21  $\pm$   0.14  &  25.42  $\pm$   0.16  &   -  &   - \\
HDF-oC37            &   2.926  &   1.36  &   0.40  &  25.25  &   -  &  24.37  $\pm$   0.48  &   -  &   - \\
HDF-oC38            &   3.110  &   1.59  &   0.67  &  24.97  &  23.39  $\pm$   0.12  &  23.23  $\pm$   0.16  &  23.64  $\pm$   0.26  &  22.79  $\pm$   0.10 \\
HDF-oD12            &   2.418  &   0.77  &   0.31  &  24.84  &   -  &   -  &   -  &   - \\
HDF-oD3             &   2.724  &   1.29  &   0.51  &  24.54  &  24.06  $\pm$   0.38  &  24.06  $\pm$   0.23  &   -  &   - \\
HDF-oMD19           &   3.241  &   1.69  &   0.75  &  24.52  &  23.27  $\pm$   0.12  &  22.91  $\pm$   0.13  &  22.85  $\pm$   0.09  &  22.28  $\pm$   0.09 \\
HDF-oMD24           &   2.942  &   1.29  &   0.34  &  24.33  &  23.63  $\pm$   0.11  &  23.57  $\pm$   0.12  &  23.76  $\pm$   0.13  &   - \\
HDF-oMD28           &   2.917  &   1.69  &   0.76  &  24.47  &  23.09  $\pm$   0.13  &  23.09  $\pm$   0.18  &  23.52  $\pm$   0.26  &  22.92  $\pm$   0.11 \\
HDF-oMD51           &   2.431  &   1.32  &   0.34  &  23.87  &  23.12  $\pm$   0.12  &  23.13  $\pm$   0.16  &  23.00  $\pm$   0.09  &  22.93  $\pm$   0.11 \\
HDF-oMD54           &   2.980  &   1.01  &   0.45  &  24.59  &  23.88  $\pm$   0.14  &  24.05  $\pm$   0.32  &  24.12  $\pm$   0.92  &  24.28  $\pm$   0.41 \\
Q1422-C101          &   2.873  &   3.79  &   0.85  &  24.17  &  21.89  $\pm$   0.09  &  21.70  $\pm$   0.09  &  21.91  $\pm$   0.13  &  21.88  $\pm$   0.10 \\
Q1422-C102          &   3.092  &   3.36  &   0.58  &  25.36  &   -  &   -  &   -  &   - \\
Q1422-C106          &   3.032  &   3.09  &   0.73  &  25.30  &  23.45  $\pm$   0.25  &  23.78  $\pm$   0.18  &   -  &   - \\
Q1422-C108          &   3.375  &   3.58  &   0.60  &  24.78  &   -  &   -  &   -  &   - \\
Q1422-C110          &   3.072  &   3.79  &   0.86  &  24.27  &  22.64  $\pm$   0.10  &  22.61  $\pm$   0.09  &   -  &  22.72  $\pm$   0.50 \\
Q1422-C118          &   2.971  &   3.35  &   0.60  &  25.14  &  23.77  $\pm$   0.19  &  23.36  $\pm$   0.17  &   -  &   - \\
Q1422-C121          &   3.748  &   2.86  &   1.03  &  25.50  &   -  &  24.19  $\pm$   0.24  &   -  &   - \\
Q1422-C42           &   3.562  &   3.25  &   1.08  &  24.37  &  21.03  $\pm$   0.07  &  20.99  $\pm$   0.06  &  20.64  $\pm$   0.09  &  20.97  $\pm$   0.08 \\
Q1422-C52           &   3.072  &   3.40  &   0.91  &  24.61  &   -  &   -  &   -  &   - \\
Q1422-C63           &   3.053  &   2.60  &   0.64  &  25.85  &   -  &  23.97  $\pm$   0.18  &   -  &   - \\
Q1422-C70           &   3.126  &   2.72  &   0.92  &  25.45  &   -  &   -  &   -  &   - \\
Q1422-C81           &   3.589  &   2.68  &   1.03  &  25.08  &  22.45  $\pm$   0.09  &  22.06  $\pm$   0.09  &   -  &   - \\
Q1422-C92           &   3.009  &   2.46  &   0.91  &  25.47  &   -  &   -  &   -  &   - \\
Q1422-C93           &   3.082  &   3.12  &   0.71  &  25.32  &  23.63  $\pm$   0.25  &  23.09  $\pm$   0.18  &  22.32  $\pm$   0.27  &  22.79  $\pm$   0.50 \\
Q1422-C99           &   3.064  &   2.84  &   0.93  &  25.14  &  23.61  $\pm$   0.25  &  23.69  $\pm$   0.16  &   -  &   - \\
Q1422-D33           &   3.074  &   3.66  &   0.86  &  24.59  &   -  &  23.65  $\pm$   0.16  &   -  &   - \\
Q1422-D42           &   3.135  &   2.69  &   0.62  &  25.32  &  23.14  $\pm$   0.39  &   -  &   -  &   - \\
Q1422-D43           &   2.970  &   2.15  &   0.23  &  25.76  &  21.04  $\pm$   0.07  &  21.05  $\pm$   0.06  &   -  &  21.51  $\pm$   0.09 \\
Q1422-D45           &   3.074  &   2.49  &   0.31  &  24.11  &  22.97  $\pm$   0.10  &  23.33  $\pm$   0.17  &  21.43  $\pm$   0.12  &   - \\
Q1422-D53           &   3.087  &   2.57  &   0.83  &  24.23  &  23.30  $\pm$   0.39  &  22.94  $\pm$   0.09  &   -  &  23.01  $\pm$   0.50 \\
Q1422-D54           &   2.938  &   2.04  &   0.52  &  25.95  &  23.11  $\pm$   0.39  &   -  &   -  &   - \\
Q1422-D63           &   2.779  &   2.00  &   0.47  &  25.29  &   -  &   -  &   -  &   - \\
Q1422-D68           &   3.290  &   2.56  &   0.39  &  24.72  &  23.51  $\pm$   0.25  &  23.30  $\pm$   0.17  &   -  &   - \\
Q1422-D72           &   3.144  &   2.70  &   0.83  &  24.86  &   -  &   -  &   -  &   - \\
Q1422-D76           &   2.939  &   3.66  &   0.38  &  24.56  &  23.67  $\pm$   0.25  &  23.24  $\pm$   0.17  &   -  &   - \\
Q1422-D77           &   2.649  &   2.59  &   0.75  &  24.31  &  20.74  $\pm$   0.05  &  20.58  $\pm$   0.05  &  20.14  $\pm$   0.09  &  19.47  $\pm$   0.05 \\
Q1422-D78           &   3.104  &   3.40  &   0.95  &  23.77  &  21.51  $\pm$   0.08  &  21.71  $\pm$   0.09  &   -  &  21.01  $\pm$   0.09 \\
Q1422-D80           &   2.913  &   3.35  &   0.15  &  24.94  &  23.40  $\pm$   0.25  &  23.45  $\pm$   0.16  &   -  &   - \\
Q1422-D81           &   3.103  &   3.53  &   0.51  &  23.41  &  21.56  $\pm$   0.08  &   -  &   -  &   - \\
Q1422-D88           &   3.755  &   2.93  &   1.20  &  24.44  &  23.07  $\pm$   0.20  &   -  &   -  &   - \\
Q1422-D91           &   2.921  &   2.34  &   0.44  &  23.67  &  22.48  $\pm$   0.09  &  22.57  $\pm$   0.09  &   -  &   - \\
Q1422-D95           &   3.227  &   2.28  &   0.62  &  25.04  &  22.85  $\pm$   0.10  &  22.70  $\pm$   0.09  &   -  &   - \\
Q1422-MD106         &   2.412  &   1.33  &   0.29  &  25.61  &   -  &   -  &   -  &   - \\
Q1422-MD111         &   2.658  &   1.57  &   0.48  &  23.44  &  21.73  $\pm$   0.09  &  21.85  $\pm$   0.09  &  21.16  $\pm$   0.09  &  22.72  $\pm$   0.50 \\
Q1422-MD119         &   3.038  &   2.04  &   0.76  &  24.99  &  23.40  $\pm$   0.25  &   -  &   -  &   - \\
Q1422-MD120         &   3.566  &   2.12  &   1.09  &  25.51  &   -  &   -  &   -  &   - \\
Q1422-MD133         &   2.747  &   1.67  &   0.26  &  23.24  &  20.99  $\pm$   0.05  &  21.12  $\pm$   0.06  &  20.66  $\pm$   0.09  &  21.12  $\pm$   0.09 \\
Q1422-MD139         &   2.746  &   2.06  &   0.91  &  24.59  &   -  &   -  &   -  &   - \\
Q1422-MD152         &   3.243  &   2.20  &   1.18  &  24.06  &  21.39  $\pm$   0.08  &  21.29  $\pm$   0.06  &  21.85  $\pm$   0.14  &  22.28  $\pm$   0.32 \\
Q1422-MD156         &   2.704  &   1.92  &   0.73  &  24.49  &  22.71  $\pm$   0.10  &  22.69  $\pm$   0.09  &   -  &  22.93  $\pm$   0.50 \\
Q1422-MD166         &   2.976  &   1.99  &   0.54  &  25.71  &   -  &   -  &   -  &   - \\
Q1422-MD172         &   2.664  &   2.32  &   1.10  &  25.18  &   -  &   -  &   -  &   - \\
Q1422-MD185         &   2.858  &   2.09  &   0.78  &  23.86  &  22.21  $\pm$   0.09  &  22.36  $\pm$   0.12  &   -  &   - \\
Q1422-MD188         &   2.560  &   2.03  &   1.01  &  25.17  &  22.51  $\pm$   0.10  &  22.43  $\pm$   0.12  &   -  &  23.98  $\pm$   0.50 \\
Q1422-MD189         &   2.914  &   2.43  &   1.01  &  24.74  &   -  &  21.83  $\pm$   0.09  &   -  &  21.56  $\pm$   0.09 \\
Q1422-MD206         &   2.787  &   1.92  &   0.81  &  24.95  &   -  &   -  &   -  &   - \\
Q1422-MD209         &   3.376  &   2.51  &   1.06  &  24.57  &   -  &   -  &   -  &   - \\
Q1422-MD213         &   2.595  &   1.96  &   0.77  &  23.72  &   -  &   -  &   -  &   - \\
Q1422-MD216         &   2.976  &   1.74  &   0.52  &  24.65  &  23.61  $\pm$   0.25  &   -  &   -  &  22.20  $\pm$   0.35 \\
Q1422-MD92          &   3.139  &   2.04  &   0.65  &  25.24  &   -  &  23.95  $\pm$   0.18  &   -  &   - \\
Q1422-MD96          &   2.853  &   1.74  &   0.45  &  25.97  &   -  &   -  &   -  &   - \\
Q1422-oC50          &   3.089  &   2.77  &   0.95  &  24.89  &  21.76  $\pm$   0.09  &  21.53  $\pm$   0.07  &  21.33  $\pm$   0.12  &  21.04  $\pm$   0.09 \\
Q2233-C10           &   3.002  &   2.31  &   0.62  &  24.88  &   -  &   -  &   -  &   - \\
Q2233-C11           &   3.110  &   3.35  &   0.77  &  23.55  &  22.02  $\pm$   0.09  &  22.45  $\pm$   0.09  &   -  &   - \\
Q2233-C12           &   3.109  &   2.53  &   0.92  &  24.39  &  22.96  $\pm$   0.09  &  22.90  $\pm$   0.10  &   -  &   - \\
Q2233-C9            &   2.874  &   2.03  &   0.47  &  25.42  &  23.42  $\pm$   0.27  &   -  &   -  &   - \\
Q2233-D4            &   2.595  &   1.77  &   0.12  &  25.22  &  22.62  $\pm$   0.09  &  22.31  $\pm$   0.09  &   -  &   - \\
Q2233-D6            &   3.064  &   2.04  &   0.44  &  24.26  &  22.18  $\pm$   0.09  &  22.29  $\pm$   0.09  &   -  &   - \\
Q2233-M10           &   3.057  &   2.42  &   1.17  &  24.16  &  22.38  $\pm$   0.09  &  22.16  $\pm$   0.12  &   -  &  21.48  $\pm$   0.08 \\
Q2233-M16           &   3.220  &   1.94  &   0.73  &  25.27  &   -  &   -  &   -  &   - \\
Q2233-M17           &   2.733  &   2.40  &   1.06  &  24.15  &  22.53  $\pm$   0.09  &  22.71  $\pm$   0.08  &   -  &   - \\
Q2233-M23           &   3.109  &   2.18  &   1.01  &  24.66  &  24.02  $\pm$   0.46  &   -  &   -  &   - \\
Q2233-MD34          &   2.169  &   1.32  &   0.26  &  25.05  &  22.82  $\pm$   0.09  &  23.00  $\pm$   0.20  &   -  &   - \\
Q2233-MD39          &   3.041  &   2.23  &   0.78  &  23.42  &  22.52  $\pm$   0.09  &  22.59  $\pm$   0.08  &   -  &   - \\
Q2233-MD41          &   2.545  &   1.56  &   0.52  &  24.56  &  22.13  $\pm$   0.09  &  22.53  $\pm$   0.08  &  23.03  $\pm$   0.50  &  21.03  $\pm$   0.09 \\
Q2233-MD44          &   2.357  &   1.44  &   0.42  &  25.43  &  24.01  $\pm$   0.46  &  24.01  $\pm$   0.32  &   -  &   - \\
Q2233-MD46          &   2.713  &   1.90  &   0.88  &  23.81  &  23.41  $\pm$   0.27  &   -  &   -  &   - \\
Q2233-MD47          &   3.105  &   2.16  &   0.82  &  25.25  &   -  &  23.20  $\pm$   0.25  &   -  &   - \\
Q2233-MD52          &   2.837  &   2.14  &   1.07  &  23.20  &  22.57  $\pm$   0.09  &  22.89  $\pm$   0.10  &   -  &   - \\
SSA22a-aug96C19     &   2.470  &   1.11  &   0.18  &  24.42  &  22.79  $\pm$   0.09  &  22.69  $\pm$   0.09  &   -  &   - \\
SSA22a-aug96C20     &   1.357  &   0.71  &   0.22  &  25.31  &  23.43  $\pm$   0.26  &  23.27  $\pm$   0.16  &   -  &   - \\
SSA22a-aug96C22     &   2.129  &   0.67  &   0.23  &  24.41  &  21.83  $\pm$   0.08  &  21.74  $\pm$   0.09  &  21.71  $\pm$   0.08  &  21.68  $\pm$   0.07 \\
SSA22a-aug96C3      &   1.674  &   1.24  &   0.34  &  24.25  &  22.74  $\pm$   0.09  &  23.10  $\pm$   0.18  &   -  &   - \\
SSA22a-aug96D11     &   0.345  &   0.55  &   0.82  &  22.79  &  22.84  $\pm$   0.10  &  23.03  $\pm$   0.18  &   -  &   - \\
SSA22a-aug96M16     &   3.292  &   1.36  &   0.70  &  23.83  &  23.56  $\pm$   0.26  &  23.66  $\pm$   0.19  &   -  &   - \\
SSA22a-aug96MD40    &   2.175  &   0.60  &   0.05  &  24.31  &  23.34  $\pm$   0.17  &  23.45  $\pm$   0.19  &   -  &   - \\
SSA22a-C10          &   2.929  &   2.21  &   0.42  &  25.08  &  23.47  $\pm$   0.26  &  23.84  $\pm$   0.20  &   -  &   - \\
SSA22a-C11          &   3.104  &   2.95  &   0.47  &  24.20  &  23.10  $\pm$   0.17  &  23.07  $\pm$   0.18  &   -  &   - \\
SSA22a-C12          &   3.112  &   2.90  &   0.44  &  24.37  &   -  &   -  &   -  &   - \\
SSA22a-C15          &   3.094  &   2.11  &   0.55  &  25.19  &   -  &   -  &   -  &   - \\
SSA22a-C16          &   3.065  &   2.88  &   0.98  &  23.64  &  21.98  $\pm$   0.09  &  21.82  $\pm$   0.09  &   -  &  21.26  $\pm$   0.09 \\
SSA22a-C22          &   2.882  &   2.72  &   0.54  &  24.46  &  23.66  $\pm$   0.26  &  23.41  $\pm$   0.19  &   -  &   - \\
SSA22a-C24          &   3.096  &   2.77  &   0.78  &  23.86  &  22.97  $\pm$   0.10  &  22.61  $\pm$   0.09  &   -  &   - \\
SSA22a-C26          &   3.178  &   2.02  &   0.34  &  25.12  &  24.17  $\pm$   0.41  &   -  &   -  &   - \\
SSA22a-C27          &   3.084  &   2.45  &   0.67  &  25.08  &  23.38  $\pm$   0.17  &  23.32  $\pm$   0.16  &   -  &   - \\
SSA22a-C28          &   3.076  &   2.52  &   0.50  &  25.08  &   -  &   -  &   -  &   - \\
SSA22a-C30          &   3.101  &   2.53  &   0.82  &  24.22  &  22.47  $\pm$   0.10  &  22.27  $\pm$   0.09  &   -  &   - \\
SSA22a-C31          &   3.021  &   3.30  &  -0.11  &  24.61  &   -  &   -  &   -  &   - \\
SSA22a-C32          &   3.296  &   3.19  &   0.67  &  23.68  &  23.34  $\pm$   0.17  &  23.34  $\pm$   0.16  &   -  &   - \\
SSA22a-C35          &   3.101  &   2.52  &   0.95  &  24.18  &  23.16  $\pm$   0.17  &  23.08  $\pm$   0.18  &   -  &   - \\
SSA22a-C36          &   3.063  &   2.86  &   0.78  &  24.06  &  22.89  $\pm$   0.10  &  22.75  $\pm$   0.09  &   -  &   - \\
SSA22a-C37          &   0.452  &   3.03  &   0.45  &  24.12  &  21.35  $\pm$   0.07  &  21.20  $\pm$   0.07  &  21.48  $\pm$   0.09  &  21.19  $\pm$   0.09 \\
SSA22a-C39          &   3.076  &   2.39  &   0.40  &  24.90  &  24.17  $\pm$   0.41  &  23.87  $\pm$   0.20  &   -  &   - \\
SSA22a-C40          &   2.922  &   2.50  &   0.37  &  25.08  &  24.64  $\pm$   0.24  &  24.23  $\pm$   0.34  &   -  &   - \\
SSA22a-C41          &   3.022  &   3.49  &   0.18  &  23.80  &  23.32  $\pm$   0.17  &  22.97  $\pm$   0.09  &   -  &   - \\
SSA22a-C42          &   2.925  &   2.69  &   0.30  &  25.10  &   -  &  24.26  $\pm$   0.34  &   -  &   - \\
SSA22a-C44          &   2.823  &   2.41  &   0.61  &  24.66  &   -  &   -  &   -  &   - \\
SSA22a-C45          &   2.826  &   2.36  &   0.49  &  24.76  &  24.30  $\pm$   0.41  &  23.77  $\pm$   0.20  &   -  &   - \\
SSA22a-C46          &   2.927  &   2.62  &   0.30  &  24.82  &  23.75  $\pm$   0.20  &   -  &   -  &   - \\
SSA22a-C48          &   3.085  &   2.56  &   0.31  &  24.71  &   -  &  24.14  $\pm$   0.34  &   -  &   - \\
SSA22a-C4           &   3.076  &   2.54  &   0.42  &  24.53  &  23.88  $\pm$   0.20  &   -  &   -  &   - \\
SSA22a-C50          &   3.086  &   2.21  &   0.58  &  25.19  &  23.41  $\pm$   0.26  &  23.30  $\pm$   0.16  &   -  &   - \\
SSA22a-C6           &   3.095  &   2.97  &   0.79  &  23.44  &  23.05  $\pm$   0.19  &  23.10  $\pm$   0.16  &   -  &   - \\
SSA22a-D14          &   3.019  &   2.29  &   0.19  &  24.32  &  23.76  $\pm$   0.20  &  23.72  $\pm$   0.20  &   -  &   - \\
SSA22a-D15          &   2.716  &   2.37  &   0.31  &  25.05  &   -  &   -  &   -  &   - \\
SSA22a-D17          &   3.089  &   2.01  &   0.45  &  24.27  &  23.47  $\pm$   0.26  &  23.50  $\pm$   0.19  &   -  &   - \\
SSA22a-D3           &   3.082  &   2.58  &   0.97  &  23.37  &  22.36  $\pm$   0.10  &  22.44  $\pm$   0.09  &   -  &   - \\
SSA22a-D4           &   2.770  &   1.72  &  -0.02  &  24.85  &   -  &   -  &   -  &   - \\
SSA22a-D7           &   2.759  &   2.14  &   0.62  &  23.50  &  22.44  $\pm$   0.10  &  22.30  $\pm$   0.09  &   -  &   - \\
SSA22a-M10          &   3.099  &   2.20  &   1.03  &  24.45  &  23.71  $\pm$   0.20  &  23.74  $\pm$   0.20  &   -  &   - \\
SSA22a-M14          &   3.091  &   1.85  &   0.75  &  25.47  &  22.12  $\pm$   0.09  &  21.75  $\pm$   0.09  &  21.82  $\pm$   0.08  &  20.90  $\pm$   0.08 \\
SSA22a-M28          &   3.091  &   2.07  &   0.82  &  24.74  &  23.06  $\pm$   0.19  &  22.98  $\pm$   0.09  &   -  &   - \\
SSA22a-M38          &   3.288  &   2.15  &   1.15  &  24.11  &  22.47  $\pm$   0.10  &  22.23  $\pm$   0.09  &  22.04  $\pm$   0.20  &   - \\
SSA22a-M4           &   3.093  &   2.22  &   0.76  &  24.83  &   -  &  23.53  $\pm$   0.19  &   -  &   - \\
SSA22a-MD14         &   3.094  &   2.25  &   0.86  &  24.14  &  23.09  $\pm$   0.19  &  23.30  $\pm$   0.16  &   -  &   - \\
SSA22a-MD17         &   2.163  &   1.56  &   0.23  &  24.82  &  22.44  $\pm$   0.10  &  22.24  $\pm$   0.09  &   -  &   - \\
SSA22a-MD19         &   2.408  &   1.57  &   0.38  &  24.62  &  22.32  $\pm$   0.10  &  22.26  $\pm$   0.09  &  21.86  $\pm$   0.08  &   - \\
SSA22a-MD20         &   3.050  &   1.58  &   0.50  &  25.28  &   -  &   -  &   -  &   - \\
SSA22a-MD23         &   3.087  &   1.92  &   0.46  &  24.14  &   -  &   -  &   -  &   - \\
SSA22a-MD2          &   2.482  &   1.64  &   0.58  &  24.66  &  23.84  $\pm$   0.20  &  23.80  $\pm$   0.20  &   -  &   - \\
SSA22a-MD32         &   3.102  &   1.45  &   0.37  &  25.41  &  24.43  $\pm$   0.18  &  24.00  $\pm$   0.20  &   -  &   - \\
SSA22a-MD36         &   2.742  &   1.94  &   0.64  &  24.76  &  23.32  $\pm$   0.17  &  23.09  $\pm$   0.18  &   -  &   - \\
SSA22a-MD37         &   3.026  &   1.76  &   0.73  &  24.73  &  23.61  $\pm$   0.26  &  23.51  $\pm$   0.19  &   -  &   - \\
SSA22a-MD3          &   2.488  &   1.30  &   0.15  &  24.60  &   -  &   -  &   -  &   - \\
SSA22a-MD40         &   3.022  &   1.78  &   0.70  &  24.89  &  23.41  $\pm$   0.26  &  23.70  $\pm$   0.20  &   -  &   - \\
SSA22a-MD41         &   2.173  &   1.31  &   0.19  &  23.31  &  22.00  $\pm$   0.09  &  21.89  $\pm$   0.09  &  22.07  $\pm$   0.20  &  23.45  $\pm$   0.50 \\
SSA22a-MD44         &   2.712  &   1.52  &   0.42  &  25.50  &   -  &   -  &   -  &   - \\
SSA22a-MD46         &   3.086  &   1.62  &   0.42  &  23.30  &  22.95  $\pm$   0.10  &  22.99  $\pm$   0.09  &   -  &   - \\
SSA22a-MD4          &   2.613  &   1.51  &   0.24  &  24.25  &  23.30  $\pm$   0.17  &  23.05  $\pm$   0.18  &   -  &   - \\
SSA22a-MD55         &   2.673  &   2.03  &   0.85  &  24.36  &  22.60  $\pm$   0.09  &   -  &  21.58  $\pm$   0.09  &   - \\
SSA22a-MD58         &   2.857  &   1.62  &   0.41  &  25.26  &  20.93  $\pm$   0.06  &  20.75  $\pm$   0.07  &  20.49  $\pm$   0.09  &  21.14  $\pm$   0.09 \\
SSA22a-MD6          &   2.582  &   1.91  &   0.89  &  24.07  &   -  &   -  &   -  &   - \\
SSA22a-oct96C1      &   2.812  &   1.26  &   0.31  &  25.32  &   -  &   -  &   -  &   - \\
SSA22a-oct96D6      &   2.852  &   1.93  &   0.04  &  25.56  &   -  &   -  &   -  &   - \\
SSA22a-oct96MD37    &   2.165  &   0.97  &  -0.01  &  24.20  &  22.83  $\pm$   0.10  &  22.88  $\pm$   0.09  &   -  &   - \\
SSA22b-C10          &   3.357  &   2.62  &   0.93  &  24.11  &  24.00  $\pm$   0.20  &   -  &   -  &   - \\
SSA22b-C12          &   3.289  &   2.59  &   0.65  &  24.94  &   -  &   -  &   -  &   - \\
SSA22b-C16          &   2.960  &   2.19  &   0.56  &  25.38  &   -  &   -  &   -  &   - \\
SSA22b-C18          &   2.684  &   2.05  &   0.51  &  24.87  &  23.69  $\pm$   0.26  &  23.75  $\pm$   0.20  &   -  &   - \\
SSA22b-C20          &   3.197  &   2.94  &   0.68  &  24.60  &  23.16  $\pm$   0.17  &  22.58  $\pm$   0.09  &   -  &   - \\
SSA22b-D10          &   2.766  &   2.46  &   0.63  &  24.32  &  23.30  $\pm$   0.17  &  23.25  $\pm$   0.16  &   -  &   - \\
SSA22b-D12          &   3.055  &   2.11  &   0.47  &  25.06  &  24.73  $\pm$   0.24  &   -  &   -  &   - \\
SSA22b-D13          &   2.627  &   2.56  &   0.62  &  23.94  &  22.94  $\pm$   0.10  &  23.34  $\pm$   0.16  &   -  &   - \\
SSA22b-D14          &   2.866  &   2.01  &   0.41  &  24.90  &   -  &  24.34  $\pm$   0.45  &   -  &   - \\
SSA22b-D15          &   2.864  &   2.08  &   0.56  &  24.65  &   -  &   -  &   -  &   - \\
SSA22b-D5           &   3.180  &   2.18  &   0.66  &  24.42  &  23.70  $\pm$   0.26  &   -  &   -  &   - \\
SSA22b-M16          &   2.956  &   2.06  &   0.79  &  25.03  &  23.32  $\pm$   0.17  &  23.21  $\pm$   0.16  &   -  &   - \\
SSA22b-M19          &   3.025  &   2.26  &   0.79  &  24.99  &   -  &  23.36  $\pm$   0.16  &   -  &   - \\
SSA22b-MD27         &   2.784  &   1.94  &   0.55  &  25.18  &   -  &   -  &  20.57  $\pm$   0.09  &   - \\
SSA22b-MD32         &   2.949  &   1.39  &   0.13  &  24.72  &   -  &   -  &   -  &   - \\
SSA22b-MD38         &   2.861  &   1.89  &   0.52  &  24.02  &  23.66  $\pm$   0.26  &  23.77  $\pm$   0.20  &   -  &   - \\
SSA22b-oC16         &   2.656  &   1.45  &   0.69  &  24.83  &  22.89  $\pm$   0.10  &  22.83  $\pm$   0.09  &   -  &  21.55  $\pm$   0.08 \\
SSA22b-oC22         &   2.962  &   1.28  &   0.56  &  24.79  &  24.08  $\pm$   0.41  &  24.27  $\pm$   0.34  &   -  &   - \\
SSA22b-oC27         &   3.370  &   1.33  &   0.60  &  24.50  &  22.34  $\pm$   0.10  &  22.87  $\pm$   0.09  &  22.84  $\pm$   0.61  &  23.57  $\pm$   0.50 \\
SSA22b-oD22         &   2.567  &   1.37  &   0.41  &  23.98  &  22.96  $\pm$   0.10  &  22.75  $\pm$   0.09  &   -  &   - \\
SSA22b-oD24         &   2.450  &   1.09  &   0.37  &  24.51  &   -  &   -  &   -  &   - \\
SSA22b-oMD58        &   2.555  &   0.91  &   0.14  &  25.18  &  24.06  $\pm$   0.41  &  23.74  $\pm$   0.20  &   -  &   - \\
SSA22b-oMD68        &   2.296  &   0.80  &   0.11  &  24.59  &  22.88  $\pm$   0.10  &   -  &   -  &   - \\

\enddata

\label{tab:sedparms}
\end{deluxetable}

\clearpage
\begin{deluxetable}{lccc}

\tabletypesize{\normalsize}
\tablewidth{0pc}
\tablecaption{CB07 best fit Stellar Population Parameters}
\tablehead{

\colhead{} &
\colhead{M$^{\ast}$} &
\colhead{Age} &
\colhead{} \\

\colhead{Name\tablenotemark{a}} &
\colhead{($10^{9}$ M$_{\odot}$)} &
\colhead{(Myr)} &
\colhead{E(B-V)}} 
\startdata

B20902-C10          &    10.394   $\pm$  0.115  &   101  &  1.566   \\
B20902-C11          &    10.384   $\pm$  0.086  &     9  &  0.589   \\
B20902-C5           &    10.943   $\pm$  0.078  &  1250  &  0.005   \\
B20902-C7           &     9.529   $\pm$  0.143  &    30  &  0.202   \\
B20902-C9           &    10.641   $\pm$  0.161  &   404  &  0.478   \\
B20902-D11          &    11.175   $\pm$  0.090  &  1434  &  0.138   \\
B20902-D14          &    10.709   $\pm$  0.136  &    71  &  1.566   \\
B20902-D8           &    10.526   $\pm$  0.101  &  1434  &  0.013   \\
B20902-M11          &    10.721   $\pm$  0.078  &   625  &  0.159   \\
B20902-MD16         &    10.549   $\pm$  0.114  &   508  &  0.689   \\
B20902-MD21         &    10.938   $\pm$  0.016  &   904  &  0.125   \\
B20902-MD25         &    10.853   $\pm$  0.077  &   806  &  0.052   \\
B20902-MD28         &    10.946   $\pm$  0.129  &  1015  &  0.138   \\
C23-Q1700           &    10.300   $\pm$  0.091  &    19  &  0.664   \\
C26-Q1700           &    10.635   $\pm$  0.091  &   255  &  0.677   \\
C7-Q1700            &    10.797   $\pm$  0.185  &  2000  &  0.337   \\
C9-Q1700            &    10.813   $\pm$  0.182  &   904  &  0.401   \\
D17-Q1700           &    11.034   $\pm$  0.056  &  2000  &  0.000   \\
D20-Q1700           &    11.008   $\pm$  0.060  &   404  &  0.098   \\
D23-Q1700           &    10.431   $\pm$  0.178  &   404  &  0.374   \\
D25-Q1700           &    10.615   $\pm$  0.113  &   508  &  0.013   \\
DSF2237b-D19        &    10.792   $\pm$  0.077  &  1015  &  0.000   \\
DSF2237b-D28        &    10.340   $\pm$  0.193  &   640  &  0.025   \\
DSF2237b-MD72       &    10.312   $\pm$  0.109  &   180  &  0.038   \\
DSF2237b-MD81       &    11.545   $\pm$  0.109  &  2000  &  0.213   \\
HDF-C14             &    11.017   $\pm$  0.182  &  1434  &  0.163   \\
HDF-C17             &    10.482   $\pm$  0.107  &   255  &  0.086   \\
HDF-C22             &     9.996   $\pm$  0.077  &    19  &  0.439   \\
HDF-C24             &     9.909   $\pm$  0.176  &    10  &  0.501   \\
HDF-C25             &    10.211   $\pm$  0.193  &   404  &  0.202   \\
HDF-C26             &    10.900   $\pm$  0.056  &   255  &  0.000   \\
HDF-C27             &    10.423   $\pm$  0.098  &   508  &  0.288   \\
HDF-C28             &    11.480   $\pm$  0.078  &  1000  &  0.265   \\
HDF-C5              &    10.158   $\pm$  0.090  &   101  &  0.938   \\
HDF-C6              &     9.834   $\pm$  0.077  &    19  &  0.439   \\
HDF-C8              &     9.676   $\pm$  0.101  &    10  &  0.353   \\
HDF-D11             &     9.748   $\pm$  0.185  &    50  &  0.077   \\
HDF-D13             &    10.881   $\pm$  0.149  &   640  &  0.276   \\
HDF-D14             &    10.119   $\pm$  0.073  &   180  &  0.138   \\
HDF-D15             &    10.271   $\pm$  0.168  &   180  &  0.163   \\
HDF-D2              &     9.701   $\pm$  0.034  &    19  &  0.551   \\
HDF-D3              &    10.487   $\pm$  0.196  &  1434  &  0.013   \\
HDF-D6              &     9.334   $\pm$  0.193  &   640  &  0.000   \\
HDF-D7              &    10.469   $\pm$  0.061  &   255  &  0.215   \\
HDF-D8              &     9.686   $\pm$  0.101  &    80  &  0.202   \\
HDF-M16             &    10.340   $\pm$  0.191  &   180  &  0.388   \\
HDF-M17             &    11.128   $\pm$  0.094  &  2000  &  0.195   \\
HDF-M18             &    11.163   $\pm$  0.067  &  1434  &  0.362   \\
HDF-M21             &    10.857   $\pm$  0.096  &  1015  &  0.276   \\
HDF-M22             &    10.390   $\pm$  0.007  &   180  &  0.013   \\
HDF-M23             &    11.460   $\pm$  0.078  &  2000  &  0.259   \\
HDF-M25             &    10.451   $\pm$  0.196  &   180  &  0.025   \\
HDF-M27             &    10.526   $\pm$  0.096  &    10  &  0.777   \\
HDF-M28             &     9.354   $\pm$  0.098  &     6  &  0.576   \\
HDF-M32             &    10.500   $\pm$  0.109  &   180  &  0.000   \\
HDF-M35             &    10.818   $\pm$  0.067  &   255  &  0.236   \\
HDF-M7              &    10.302   $\pm$  0.094  &    10  &  0.739   \\
HDF-M9              &    10.840   $\pm$  0.097  &   404  &  0.125   \\
HDF-MD10            &     9.887   $\pm$  0.101  &   180  &  0.086   \\
HDF-MD18            &     9.454   $\pm$  0.143  &   101  &  0.150   \\
HDF-MD19            &    10.396   $\pm$  0.103  &   806  &  0.098   \\
HDF-MD22            &     9.893   $\pm$  0.049  &    10  &  0.465   \\
HDF-MD37            &    10.026   $\pm$  0.189  &   255  &  0.215   \\
HDF-MD3             &     9.887   $\pm$  0.078  &   360  &  0.000   \\
HDF-MD40            &    10.538   $\pm$  0.139  &   806  &  1.253   \\
HDF-oC26            &     8.992   $\pm$  0.082  &   180  &  0.052   \\
HDF-oC29            &     9.113   $\pm$  0.091  &     9  &  0.465   \\
HDF-oC37            &     9.333   $\pm$  0.193  &   806  &  0.000   \\
HDF-oC38            &    10.511   $\pm$  0.141  &   404  &  0.013   \\
HDF-oD3             &     9.754   $\pm$  0.099  &    90  &  0.276   \\
HDF-oMD19           &    10.797   $\pm$  0.168  &  1800  &  0.077   \\
HDF-oMD24           &    10.109   $\pm$  0.101  &   404  &  0.000   \\
HDF-oMD28           &    10.533   $\pm$  0.191  &  1278  &  0.077   \\
HDF-oMD51           &     9.884   $\pm$  0.129  &   180  &  0.190   \\
HDF-oMD54           &     9.693   $\pm$  0.073  &   255  &  0.000   \\
Q1422-C101          &    11.007   $\pm$  0.159  &   604  &  0.865   \\
Q1422-C106          &    10.433   $\pm$  0.193  &   640  &  0.478   \\
Q1422-C110          &    10.813   $\pm$  0.014  &   404  &  0.664   \\
Q1422-C118          &    10.340   $\pm$  0.075  &   180  &  0.637   \\
Q1422-C121          &     9.857   $\pm$  0.054  &    10  &  0.714   \\
Q1422-C42           &    11.925   $\pm$  0.098  &  1015  &  0.487   \\
Q1422-C63           &    10.211   $\pm$  0.084  &  2000  &  0.251   \\
Q1422-C81           &    11.289   $\pm$  0.155  &  1278  &  0.301   \\
Q1422-C93           &    10.755   $\pm$  0.189  &  1434  &  0.465   \\
Q1422-C99           &    10.340   $\pm$  0.193  &   640  &  0.426   \\
Q1422-D33           &    10.037   $\pm$  0.094  &   101  &  0.499   \\
Q1422-D42           &    10.566   $\pm$  0.116  &  1015  &  0.202   \\
Q1422-D45           &    10.943   $\pm$  0.078  &  2000  &  0.050   \\
Q1422-D53           &    10.362   $\pm$  0.098  &   180  &  0.000   \\
Q1422-D54           &    10.777   $\pm$  0.107  &  1434  &  0.038   \\
Q1422-D68           &    10.618   $\pm$  0.103  &   806  &  0.138   \\
Q1422-D76           &     9.873   $\pm$  0.195  &    19  &  0.702   \\
Q1422-D77           &    11.751   $\pm$  0.078  &  2000  &  0.874   \\
Q1422-D78           &    11.398   $\pm$  0.058  &  1278  &  0.465   \\
Q1422-D80           &     9.989   $\pm$  0.076  &    71  &  0.677   \\
Q1422-D81           &    11.304   $\pm$  0.099  &  1278  &  0.362   \\
Q1422-D88           &    11.037   $\pm$  0.133  &   508  &  0.301   \\
Q1422-D91           &    10.681   $\pm$  0.065  &   404  &  0.263   \\
Q1422-D95           &    10.825   $\pm$  0.074  &  1015  &  0.113   \\
Q1422-MD111         &    10.998   $\pm$  0.109  &  1434  &  0.276   \\
Q1422-MD119         &    10.706   $\pm$  0.074  &  1015  &  0.086   \\
Q1422-MD133         &    11.321   $\pm$  0.056  &  1015  &  0.413   \\
Q1422-MD152         &    11.379   $\pm$  0.060  &  1015  &  0.175   \\
Q1422-MD156         &    10.521   $\pm$  0.133  &   404  &  0.677   \\
Q1422-MD185         &    10.918   $\pm$  0.161  &   806  &  0.413   \\
Q1422-MD188         &    10.543   $\pm$  0.135  &   255  &  0.689   \\
Q1422-MD189         &    11.226   $\pm$  0.109  &  2000  &  0.224   \\
Q1422-MD216         &    10.575   $\pm$  0.101  &  1278  &  0.077   \\
Q1422-MD92          &    10.211   $\pm$  0.168  &  1800  &  0.013   \\
Q1422-oC50          &    10.957   $\pm$  0.143  &  1434  &  0.238   \\
Q2233-C11           &    10.703   $\pm$  0.092  &    19  &  0.877   \\
Q2233-C12           &    10.806   $\pm$  0.113  &   508  &  0.313   \\
Q2233-C9            &    10.698   $\pm$  0.032  &   904  &  0.426   \\
Q2233-D4            &    10.764   $\pm$  0.189  &  1800  &  0.702   \\
Q2233-D6            &    11.093   $\pm$  0.143  &  1015  &  0.052   \\
Q2233-M10           &    11.176   $\pm$  0.098  &  1434  &  0.374   \\
Q2233-M17           &    10.643   $\pm$  0.098  &    90  &  0.840   \\
Q2233-M23           &    10.300   $\pm$  0.043  &     9  &  0.702   \\
Q2233-MD34          &    10.088   $\pm$  0.109  &   255  &  0.086   \\
Q2233-MD39          &    10.721   $\pm$  0.191  &   640  &  0.113   \\
Q2233-MD41          &    10.965   $\pm$  0.078  &  1500  &  0.455   \\
Q2233-MD44          &     9.803   $\pm$  0.101  &    71  &  0.827   \\
Q2233-MD46          &    10.239   $\pm$  0.117  &     9  &  0.763   \\
Q2233-MD47          &    10.530   $\pm$  0.084  &   404  &  0.013   \\
Q2233-MD52          &    10.404   $\pm$  0.085  &   101  &  0.376   \\
SSA22a-aug96C19     &    10.396   $\pm$  0.172  &   508  &  0.512   \\
SSA22a-aug96C20     &    10.512   $\pm$  0.058  &    19  &  1.878   \\
SSA22a-aug96C22     &     9.889   $\pm$  0.058  &    19  &  1.878   \\
SSA22a-aug96C3      &     9.978   $\pm$  0.116  &    90  &  1.566   \\
SSA22a-aug96D11     &    10.204   $\pm$  0.069  &   180  &  0.689   \\
SSA22a-aug96M16     &    10.232   $\pm$  0.102  &   101  &  0.052   \\
SSA22a-aug96MD40    &    10.119   $\pm$  0.191  &   640  &  0.038   \\
SSA22a-C10          &    10.238   $\pm$  0.092  &   255  &  0.025   \\
SSA22a-C11          &    10.480   $\pm$  0.105  &   180  &  0.276   \\
SSA22a-C16          &    11.274   $\pm$  0.174  &  1434  &  0.426   \\
SSA22a-C22          &    10.157   $\pm$  0.005  &    64  &  0.628   \\
SSA22a-C24          &    10.651   $\pm$  0.075  &   255  &  0.388   \\
SSA22a-C26          &     9.923   $\pm$  0.049  &   255  &  0.013   \\
SSA22a-C27          &    10.575   $\pm$  0.101  &  1015  &  0.328   \\
SSA22a-C30          &    11.024   $\pm$  0.077  &   806  &  0.288   \\
SSA22a-C32          &    10.168   $\pm$  0.084  &    40  &  0.098   \\
SSA22a-C35          &    10.570   $\pm$  0.105  &   180  &  0.301   \\
SSA22a-C36          &    10.682   $\pm$  0.071  &   180  &  0.362   \\
SSA22a-C37          &    11.480   $\pm$  0.078  &  1000  &  3.496   \\
SSA22a-C39          &    10.211   $\pm$  0.193  &   255  &  0.224   \\
SSA22a-C40          &     9.759   $\pm$  0.094  &    19  &  0.589   \\
SSA22a-C41          &    10.047   $\pm$  0.101  &    30  &  0.301   \\
SSA22a-C42          &    10.118   $\pm$  0.094  &    19  &  0.526   \\
SSA22a-C45          &     9.531   $\pm$  0.098  &    30  &  0.702   \\
SSA22a-C46          &    10.384   $\pm$  0.191  &   180  &  0.163   \\
SSA22a-C48          &    10.076   $\pm$  0.089  &   101  &  0.150   \\
SSA22a-C4           &    10.129   $\pm$  0.066  &    80  &  0.215   \\
SSA22a-C50          &    10.555   $\pm$  0.143  &   404  &  0.013   \\
SSA22a-C6           &    10.126   $\pm$  0.049  &    30  &  0.263   \\
SSA22a-D14          &    10.109   $\pm$  0.193  &   404  &  0.025   \\
SSA22a-D17          &    10.340   $\pm$  0.193  &   640  &  0.013   \\
SSA22a-D3           &    10.883   $\pm$  0.069  &   508  &  0.215   \\
SSA22a-D7           &    10.510   $\pm$  0.123  &    64  &  0.877   \\
SSA22a-M10          &     9.784   $\pm$  0.096  &     9  &  0.639   \\
SSA22a-M14          &    11.281   $\pm$  0.243  &  1800  &  0.061   \\
SSA22a-M28          &    10.664   $\pm$  0.101  &  1434  &  0.224   \\
SSA22a-M38          &    11.097   $\pm$  0.069  &  1015  &  0.313   \\
SSA22a-M4           &    10.442   $\pm$  0.193  &   640  &  0.263   \\
SSA22a-MD14         &    10.526   $\pm$  0.101  &  1015  &  0.077   \\
SSA22a-MD17         &    10.192   $\pm$  0.135  &   255  &  0.301   \\
SSA22a-MD19         &    10.445   $\pm$  0.060  &   255  &  0.752   \\
SSA22a-MD2          &     9.679   $\pm$  0.107  &    50  &  0.439   \\
SSA22a-MD32         &    10.119   $\pm$  0.189  &  1278  &  0.000   \\
SSA22a-MD36         &    10.252   $\pm$  0.084  &   180  &  0.689   \\
SSA22a-MD37         &    10.442   $\pm$  0.191  &   640  &  0.138   \\
SSA22a-MD40         &    10.427   $\pm$  0.196  &   255  &  0.013   \\
SSA22a-MD41         &    10.348   $\pm$  0.172  &   255  &  0.138   \\
SSA22a-MD46         &    10.350   $\pm$  0.067  &   255  &  0.000   \\
SSA22a-MD4          &    10.321   $\pm$  0.161  &   180  &  0.163   \\
SSA22a-MD55         &    10.919   $\pm$  0.041  &    30  &  1.878   \\
SSA22a-MD58         &    11.140   $\pm$  0.988  &  3000  &  0.175   \\
SSA22a-oct96MD37    &     9.990   $\pm$  0.109  &   180  &  0.000   \\
SSA22b-C10          &     9.926   $\pm$  0.147  &    50  &  0.362   \\
SSA22b-C18          &    10.058   $\pm$  0.166  &    50  &  0.788   \\
SSA22b-C20          &    10.859   $\pm$  0.077  &   806  &  0.328   \\
SSA22b-D10          &    10.264   $\pm$  0.036  &   101  &  0.650   \\
SSA22b-D12          &     9.748   $\pm$  0.102  &    10  &  0.413   \\
SSA22b-D13          &    10.240   $\pm$  0.025  &   101  &  0.564   \\
SSA22b-D14          &     9.863   $\pm$  0.094  &    90  &  0.263   \\
SSA22b-D5           &    10.340   $\pm$  0.191  &   640  &  0.098   \\
SSA22b-M16          &    10.533   $\pm$  0.101  &  1015  &  0.276   \\
SSA22b-M19          &    10.575   $\pm$  0.103  &   904  &  0.313   \\
SSA22b-MD38         &     9.753   $\pm$  0.065  &    71  &  0.288   \\
SSA22b-oC16         &    10.797   $\pm$  0.078  &  2000  &  0.243   \\
SSA22b-oC22         &     9.887   $\pm$  0.101  &   180  &  0.098   \\
SSA22b-oC27         &    10.841   $\pm$  0.097  &   904  &  0.025   \\
SSA22b-oD22         &    10.212   $\pm$  0.111  &    40  &  0.865   \\
SSA22b-oMD58        &     9.767   $\pm$  0.021  &   255  &  0.077   \\
SSA22b-oMD68        &    10.396   $\pm$  0.189  &  1434  &  0.175   \\

\enddata

\tablenotetext{a}{We did not fit the stellar populations of galaxies
that had no data longward of R-band, had uncertain redshifts,
or are identified as AGN/QSO from their optical spectra. We also do not present
SED parameters for those galaxies with optical and IRAC photometry inconsistent
with a simple stellar population (these sources had large $\chi^2 > 10$ }
\label{tab:sedparms}

\end{deluxetable}
\clearpage
\newpage

\end{document}